\documentclass[onefignum,onetabnum]{siamart190516}

\ifpdf
\hypersetup{
  pdftitle={$k$-hop Steiner Tree in Tree-Like Metrics},
  pdfauthor={M. B\"ohm, R. Hoeksma, N. Megow, L. N\"olke and B. Simon}
}
\fi

\usepackage{lipsum}
\usepackage{amsfonts}
\usepackage{graphicx}
\usepackage{epstopdf}
\ifpdf
  \DeclareGraphicsExtensions{.eps,.pdf,.png,.jpg}
\else
  \DeclareGraphicsExtensions{.eps}
\fi

\newsiamremark{remark}{Remark}
\newsiamremark{hypothesis}{Hypothesis}
\crefname{hypothesis}{Hypothesis}{Hypotheses}
\newtheorem*{claim}{Claim}
\newsiamthm{observation}{Observation}
\headers{$k$-hop Steiner Tree in Tree-Like Metrics}{M. B\"ohm, R. Hoeksma, N. Megow, L. N\"olke and B. Simon}%

\title{%
On Hop-Constrained Steiner Trees in Tree-Like Metrics\thanks{{Parts of the results were} presented at MFCS 2020\cite{mfcs}.
\funding{Partially supported by the German Science Foundation (DFG) under contracts ME 3825/1 and 146371743 -- TRR 89 Invasive Computing.} }} 

\author{Martin B\"ohm\thanks{Institute of Computer Science, University of Wroc{\l}aw, Poland
  (\email{boehm@cs.uni.wroc.pl}).}
\and Ruben Hoeksma\thanks{Department of Applied Mathematics, University of Twente, The Netherlands (\email{r.p.hoeksma@utwente.nl}).}
\and Nicole Megow\thanks{Department of Mathematics and Computer Science, University of Bremen, Germany
  (\email{nicole.megow@uni-bremen.de}, \email{noelke@uni-bremen.de}).}
\and Lukas N\"olke\footnotemark[4]
\and Bertrand~Simon\thanks{IN2P3 Computing Center, CNRS, Villeurbanne, France (\email{bertrand.simon@cc.in2p3.fr}).}
}

\usepackage{amsopn}

\usepackage[utf8]{inputenc}
\usepackage[T1]{fontenc}
\usepackage{hyperref}

\usepackage{amsmath,amsfonts,bm,mathtools,amssymb} %

\usepackage{cleveref}
\usepackage{xspace}
\usepackage{geometry}
\usepackage{cite}

\usepackage{thmtools}
\usepackage{thm-restate}

\usepackage{paralist}
\usepackage{enumitem}
\usepackage{wrapfig}
\usepackage{url}
\colorlet{utgreen}{green!50!black}
\colorlet{utred}{red!50!black}

\usepackage{tikz}
\usetikzlibrary{arrows,arrows.meta,math,patterns,decorations, shapes}
\usetikzlibrary{calc, fit, shapes,intersections}
\usetikzlibrary{decorations,decorations.pathreplacing,decorations.pathmorphing,patterns, calligraphy}
\usetikzlibrary{positioning}

\tikzstyle{vertex} = [fill=black,circle,scale=0.5]
\pgfdeclarelayer{bg}    %
\pgfsetlayers{bg,main}  %

\newcommand{\mc}[1]{\ensuremath{\mathcal{#1}}\xspace}
\newcommand{\dist}[1]{\ensuremath{\mathrm{d}(#1)}\xspace}
\newcommand{\dis}{\ensuremath{\mathrm{d}}\xspace}
\newcommand{\lab}{\ensuremath{\mathrm{\ell}}\xspace}
\newcommand{\lap}{LAP\xspace}
\newcommand{\laps}{LAPs\xspace}
\newcommand{\opt}{\ensuremath{\mathrm{OPT}}\xspace}
\newcommand{\clo}{\ensuremath{\mathrm{closest}}\xspace}
\newcommand{\anch}{\ensuremath{\alpha}\xspace}
\newcommand{\Term}{\ensuremath{\mc X}\xspace}
\newcommand{\kMST}{$k$-hop MST\xspace} %
\newcommand{\kST}{$k$-hop M\smash{\v S}T\xspace} %
\newcommand{\Pol}{\ensuremath{\textrm{P}}}
\newcommand{\NP}{\ensuremath{\textrm{NP}}}
\newcommand{\diameter}{d}

\newcommand{\emptylabel}{\infty}
\newcommand{\emptynode}{v_{\emptyset}} %
\newcommand{\topt}{\opt}
\newcommand{\stree}{\ensuremath{\check{S}}\xspace}

\newcommand{\Nat}{\mathbb{N}}

\tikzstyle{vertex} = [draw=black,fill=white,circle,scale=0.5]
\tikzstyle{svertex} = [draw=utgreen,fill=black,circle,scale=0.55, very thick]
\tikzstyle{terminal} = [draw=utgreen,fill=utgreen,circle,scale=0.7]
\tikzstyle{root} = [draw=utgreen,fill=utgreen,rectangle,scale=1]
\tikzstyle{Steiner} = [draw=utgreen, ultra thick]

\usepackage[disable]{todonotes}

\newcommand{\optk}{\ensuremath{\mathit{OPT_k}}\xspace}

\begin{document}

\maketitle

\begin{abstract}
		We consider the problem of computing a Steiner tree of minimum cost under a
		hop constraint which requires the depth of the tree to be at most $k$. Our main result is an exact algorithm for metrics  induced by graphs with bounded treewidth that runs in time $n^{O(k)}$. For the special case of a path, we give a simple algorithm that solves the problem in polynomial time, even if $k$ is part of the input. The main result can be used to obtain, in quasi-polynomial time, a near-optimal solution that violates the $k$-hop constraint by at most one hop for more general metrics induced by graphs of bounded highway dimension and bounded doubling dimension. For non-metric graphs, we rule out an $o(\log n)$-approximation, assuming P$\neq$NP even when relaxing the hop constraint by any additive constant.

	\end{abstract}

\begin{keywords}
  k-hop Steiner tree, hop-constrained, dynamic programming, bounded treewidth
\end{keywords}

\begin{AMS}
  68Q25, 90C27, 05C12  
\end{AMS}

\section{Introduction}
       	\label{sec:Intro}
 
        The \emph{minimum-cost Steiner tree} problem {is a fundamental network design problem: Given} %
        a set of terminals in a finite metric space, {the task is} %
        to find a tree that spans all terminals and has minimum overall cost. {While }%
        present %
        in the original {list of Karp's} 21 NP-complete problems\cite{karp21}, %
        research on this problem goes {back as far as} the 1930s~\cite{JarnikKoessler}, and its origins can be traced back even
        further, as it is named after the 19th century Swiss mathematician Jakob
        Steiner.
        {When} the set of terminals to be interconnected is equal to the ground
        set of the metric space, we speak of the equally fundamental \emph{minimum-cost spanning tree problem},
        which is solvable {in polynomial time}. Both problems have had major applications in the 20th and 21st century,
        particularly in the design of transportation and communication networks. For {a comprehensive discussion of the} background on these two problems,
        we refer to the book~\cite{steiner-book} and the historical treatise~\cite{mst-history}. 

        When considering the efficiency and reliability of a %
        network, %
        {it is a common} {and natural} requirement that {vertices are not just {simply connected{,} but rather connected with a path that consists of} only few edges}.
        In the literature on network design, this requirement is known as \emph{bounded hop distance}, where
        \emph{hop} refers to {an edge and \emph{hop distance} to the number of edges on a path}. %
        A restriction on hop distances aims at {reducing transmission delays and packet loss}, {avoiding} the flooding of a network when routing, and increasing reliability of networks 
	by limiting the amplifying effect of link failures. There exists a multitude of applications; see, e.g., \cite{CarmiCT18,Voss99,BalakrishnanA92,Gouveia1995,DahlGR06,GouveiaSU11,Haenggi2004,Saksena1989}.

	Despite their {practical} relevance, adding hop constraints makes network design problems substantially harder. The minimum-cost spanning tree problem, for example, is well-known to be polynomial-time solvable, whereas its hop-constrained variants {do admit constant lower bounds on the approximation ratio~\cite{Guha1999, Alfandari1999} in certain metrics.}
	{However, network design problems (without hop constraints) often become easier when the underlying metric can be represented as a tree or is somewhat close to a tree such as graphs {with} bounded treewidth. In this paper, we investigate whether this is the case also in presence of hop constraints.}

    We now formally define the Steiner tree problem with %
    bounded hop distance. Here, we are given a finite metric space $(V,\dis)$ with a set $V$ of $n$ points as well as a distance function $\dis: V\times V \rightarrow \mathbb{Q}_+$, a set of terminals $\Term\subseteq V$, a root $r\in \Term$, and an integer~$k\geq 1$. A~{\em k-hop Steiner tree} is {said to be} a tree $\stree=(V_{\stree},E_{\stree})$ rooted at $r$ that spans all points in~$\Term$
	and has a depth of at most~$k$. That is, $\Term\subseteq V_{\stree} \subseteq  V$ and for $v\in V_{\stree}$, the number of edges in the $r$-$v$ path in~$\stree$ is at most~$k$. The \emph{cost} of a Steiner tree refers to the sum of its edge costs, $\sum_{\{u,v\}\in E_{\stree}}\dist{u,v}$, with edge costs given by $\dis$. We consider the {\em minimum-cost k-hop Steiner tree problem}~(\kST problem\footnote{For brevity and as {an} homage to the work of Jarník and Kössler~\cite{JarnikKoessler,jarniksteiner},
    we use the Czech letter {\v S} to distinguish Steiner trees from spanning trees  in M\v ST resp.~MST. The pronunciation of {\v S} is $\langle\,$sh$\,\rangle$, the same as the German pronunciation of the letter S in Steiner.})
	that asks for a $k$-hop Steiner tree of minimum cost. 
	When $\Term = V$, this problem
	is equivalent to the {\em minimum-cost k-hop spanning tree (\kMST) problem}.

	In this work, we show how to solve the \kST problem in certain tree-like metrics. That is, we consider
	metrics which are represented by graphs from certain tree-like graph classes using the natural correspondence between metric spaces and weighted complete graphs via the shortest path metric. 
	We say a weighted graph $G=(V,E)$ {with a weight function~$d:E\rightarrow\mathbb{R}^+$}
	{\em induces} a metric $(V,\dis)$ if for any two vertices $u,v\in V$ the length of the shortest $u$-$v$ path in $G$ equals $\dist{u,v}$. 
	{While, in this paper, we do differentiate results based on the structure of the graph~$G$, 
	we do not do so based on the weight function~$d$. Therefore, whenever a graph~$G$ is said to induce 
	a metric, this implies that~$G$ has an accompanying arbitrary weight function~$d$.}
	A~metric is called a {\em tree (resp.\ path) metric} if there is a tree (resp.\ path) inducing it, and it is called a {\em metric {with} bounded treewidth} if it is induced by some graph with bounded treewidth. For a given metric, it can be decided in polynomial time if it is a path metric, a tree metric, or a
	metric {with} constant treewidth $\omega$; 
	details are outlined in Section~\ref{sec:Prelim}.
	For convenience,
	we may not always
	distinguish between a metric and the graph inducing it.
	
	{A particular tree metric was studied by Althaus et al.~\cite{AlthausFHKRS05}. They design an optimal algorithm tailored to  metrics that {can be} represented as a {\em hierarchically separated tree (HST)}. An HST is a tree with a very regular cost structure, where the {costs} of the edges on any root-to-leave path are geometrically decreasing, and where the points of the metric space appear as leaves. In such a structure, subtrees of a solution tree can be described by subintervals of leaf indices and admit a dynamic program.}
	{However, general tree metrics exhibit a much more complex structure and inner nodes of the tree must be handled with extreme care.}

	{{In this context, probabilistic tree embedding is a celebrated tool that forms} the basis of many algorithms for network design problems in arbitrary metric spaces. 
	{Any} metric can be approximated within a logarithmic factor by a distribution over trees, as was shown by Fakcharoenphol, Rao and Talwar~\cite{FakcharoenpholRT04}.
	It is a common approach of network design algorithms to {first} embed the given metric probabilistically in a tree, {then} solve the actual problem on {this} tree (optimally or approximately), and {finally} project the solution back into the original metric. In fact, this is the approach of Althaus et al.~\cite{AlthausFHKRS05} to obtain an $O(\log n)$-approximation algorithm for \kST in general metrics. Notice that it is inherent to the approach that the approximation factor cannot be better than logarithmic. While the mentioned {embedding schemes} are not capable of preserving hop constraints, very recently, Haeupler, Hershkowitz and Zuzic~\cite{HaeuplerHZ2020} proposed a new framework for approximating hop-contrained distances with partial tree metrics. It allows to reduce the hop-constrained problem to the problem without hop constraints in a tree and admits polylogarithmic bicriteria-approximations with respect to both, cost and hop distance. However, exact algorithms, constant-factor and/or uni-criteria approximation results seem out of reach for such methods.} 

	\subsection{Further Related Work}
	Hop-constrained problems have been studied since the 1980s.
	Various well-studied problems are in fact special cases of the \kST problem,
	most notably, the \kMST problem, where $\Term = V$, the Minimum Steiner Tree problem, where $k\ge n-1$, and the Uncapacitated Facility Location problem, where $k=2$. Hardness and inapproximability results {for any of these problems} 
	are therefore valid for \kST as well. In particular, \kST is \NP-hard~\cite{AlthausFHKRS05}, even for graph metrics, while the Minimum Steiner Tree problem is polynomial-time solvable on graphs {with} bounded treewidth~\cite{chimani2012improved}. {It shall be mentioned that the Minimum Steiner Tree problem in arbitrary metrics can be solved by the classical dynamic programming-based Dreyfus-Wagner algorithm~\cite{DreyfusW71} with running time $O(3^{t} \cdot n)$, where $t=|\Term|$; via fast subset convolution the running time can be improved to $O(2^t \cdot n)$ \cite{BjorklundHKK07}.}

        {When considering {\kST in} general metric spaces as well as non-metric distance functions, several hardness results are known.}
	Bern and Plassmann~\cite{Bern1989} show that the Steiner tree problem on a metric induced by a complete graph with edge weights $1$ or $2$ is MaxSNP-hard. The same is shown for metric 2-hop MST by Alfandari and Paschos~\cite{Alfandari1999}. Thus, these problems do not admit a PTAS, unless $\Pol=\NP$~\cite{AroraLMSS1998}.
	{For general non-metric distance functions {defined} by a graph,} Manyem and Stallmann~\cite{Manyem1996} show that \kST on a graph with unit-weight edges
	and $2$-hop MST cannot admit a constant-factor approximation algorithm. They also show that $k$-hop MST on a graph with edge weights $1$ or $2$ cannot admit a PTAS.
	{When weights on the graph edges are unconstrained,} Alfandari and Paschos~\cite{Alfandari1999} prove that even for 2-hop MST no~$(1-\varepsilon)\log(n)$-approximation can exist unless $\NP \subseteq \text{DTIME}[n^{O(\log\log n)}]$.
	
	The following works, %
        conceptually closest to our paper, focus on
	approximation algorithms. \mbox{Kortsarz} and Peleg~\cite{Kortsarz1997}
	consider \kST on non-metric graphs obtaining an
	approximation factor $O(\log n)$ for constant $k$ and~$O(n^\varepsilon)$ otherwise.  Althaus et al.~\cite{AlthausFHKRS05}
	give an~$O(\log n)$-approximation for metric \kMST for arbitrary $k$ that first uses
	a randomized embedding of the given metric into a
	hierarchically-separated tree %
	and then solves this problem
	optimally. %
	For	constant $k$, Laue and Matijevi\'c~\cite{LaueM08} derive a
	PTAS for \kST in the Euclidean plane.
	Their algorithm implies a
	QPTAS for Euclidean spaces of higher dimensions.
	While the first constant-factor approximation algorithm for metric \kST is due to Kantor and Peleg~\cite{KantorP09}, the attained approximation
	factor $1.52\cdot 9^{k-2}$ is prohibitively high.
	For $k=2$, a nearly optimal
	algorithm is known{: the} best known approximation ratio of $1.488$ for metric Uncapacitated Facility Location~\cite{Li2013} and
	lower bound of $1.463$~\cite{Guha1999} are valid for metric $2$-hop MST as well.

	The bounded-diameter minimum Steiner tree problem~\cite{KantorP09,GouveiaSU11} is also closely related to {the} bounded-hop problem, yet neither a generalization, nor a special case. Here, for given~$\diameter$ we look for a minimum-cost Steiner tree with diameter at most $\diameter$.
	For constant~$\diameter$, an~$O(1)$-approximation algorithm is known for graph metrics~\cite{KantorP09}. For non-metric cost functions, an $o(\log n)$-approximation algorithm has been ruled out, assuming $\Pol\neq\NP$~\cite{Bar-IlanKP01}.
	
	Furthermore, shallow-light and buy-at-bulk Steiner trees~\cite{Arya1995,Elkin2011,Chimani2015,HajiaghayiKS09,Kortsarz1997} are conceptually similar to $k$-hop M\v STs. However, a key difference is that, here, lengths of paths
	in the tree are bounded w.r.t.\ metric distance instead of the number of edges on the path. Elkin and Solomon~\cite{Elkin2011} additionally bound the number of hops, but do so by 
	$O(\log n)$ to
	bound other measures of interest. 
	Chimani and Spoerhase~\cite{Chimani2015} consider two different measures for distance and 
	weight
	and achieve an $n^\varepsilon$-approximation, violating 
	the distance by a factor~of~$1+\varepsilon$.
	
	Minimum-cost $k$-hop spanning and Steiner trees have been studied in the
	context of random graphs as well. There, the goal is to give estimates on the weight of an
	optimal tree. In this setting, sharp {thresholds} for $k$ are known~\cite{AngelFW12}.

	\subsection{Our Results}\label{sec:results}
	{We give polynomial-time algorithms that optimally solve the \kST problem in certain tree-like metrics. Our main and most general result is a dynamic program (DP) for metrics with treewidth $\omega$ with a running time of~$n^{O(\omega k)}$. As stepping stones towards this result, we first present some key techniques by considering algorithms for simpler metrics, namely the path metric and tree metric. Later, we show how to utilize our general algorithm for (bicriteria) approximation algorithms for other metric spaces.}
	
	In \Cref{sec:DPline}, we give a quite simple exact algorithm for \kST on the path metric with running time $O(kn^5)$, where $n$ is the
        number of vertices and $k$ the number of hops. Thus, the algorithm retains the polynomial running time even when $k$ is part of the input. 

        In \Cref{sec:DPtrees}, we consider the special case of tree metrics, {i.e., graphs {with} treewidth one,
        establishing the essential building blocks for our general algorithm in a more accessible setting.}
        The running time of our algorithm for tree metrics is $n^{O(k)}$.

        Let us give a few high-level insights for our algorithm on tree metrics. A tree metric naturally lends itself to recursive computation based on the structure {of the underlying tree}, so we compute an optimal {partial} solution for a sub-tree %
        rooted at some vertex $v$ (denoted by $T[v]$) {by using the solutions of its children}. 
        We index a dynamic programming cell by $v$ {and beyond that---as more information is required on how to connect the subtree to the remainder of the tree---}by $2k$ additional vertices which represent possible parents of~$v$ at different depths in a~\kST. 
        {We refer to these $2k$ additional vertices as \emph{anchoring guarantees}}
        and for each possible depth in the $k$-hop Steiner tree,
        there is one anchoring guarantee inside of~$T[v]$ and one outside.
        {The crucial decision that the DP needs to make is how to correctly propagate these anchoring guarantees when advancing in the recursion.}

 	{Finally,} in \Cref{sec:DPboundedTW}, we present {our main result,} the general version of our dynamic program that applies to all graph metrics {with} bounded treewidth.
        Specifically, our algorithm computes the optimal \kST for metrics {with} treewidth $\omega$ in time $n^{O(\omega k)}$. {To this end, we consider a so-called \emph{nice tree decomposition} of the {input graph}, a specifically structured tree whose nodes are \emph{bags} containing at most $\omega+1$ {input vertices}, and index our DP cells by a bag $b$ and the $2k$ possible parents, i.e. anchoring guarantees, of each vertex $v$ in the bag.
        This determines the DP table size and runtime of $n^{O(\omega k)}$.}
       
        Our dynamic program for tree metrics is substantially different from {the aforementioned DP by Althaus et al.}~\cite{AlthausFHKRS05} that is tailored to {HSTs.} %
        {Further, }while the DP for %
        {the plane, the two-dimensional Euclidean space, by Laue and Matijevi{\'c}}~\cite{LaueM08} has similarities to our construction for tree
	metrics, a notable difference lies in the indexing of their cells
	by distances. In our case, such a strategy does not carry enough information; hence,
	we resort to indexing by vertices, as explained above, and retain more structure.

	Our general algorithm also facilitates a quasi-polynomial time approximation algorithm for more general metrics induced by graphs of bounded highway dimension.
        This graph class was introduced by {Abraham et al.}~\cite{highwayintro} to model transportation networks.
	Intuitively, in graphs of bounded highway dimension, locally, there exists a small set of transit vertices
	such that the shortest paths between two distant vertices
	pass through some transit vertex. We provide the full definition and all details in \Cref{sec:highway}.
	
	Using a framework {by Feldmann et al.}~\cite{highway}, we show that our approach for bounded treewidth metrics and the constant-factor approximation {by Peleg and Kantor}~\cite{KantorP09} can be combined
        to {design} an algorithm that computes, in quasi-polynomial time, a $(k+1)$-hop Steiner tree of cost at most $(1+\varepsilon)\optk$, where \optk is the cost
        of the (minimal) \kST. This seems to be the first result
        {taking the advantage of a slight relaxation of the hard hop constraints in network design,}
	a research direction %
	proposed {by Althaus et al.}~\cite{AlthausFHKRS05}.
	
	Additionally, we consider the concept of doubling dimension {that was proposed by} Gupta, Krauthgamer and Lee~\cite{GuptaKL03}. A metric space is said to have doubling dimension $d$ if every ball of radius $2r$ can be covered by $2d$ balls of radius~$r$.
	While this concept is closely related to highway dimension {and Abraham et al.~\cite{AbrahamDFGW16} %
	show that  constant highway dimension implies constant doubling dimension, they also show that the converse does not hold.} %
      However, we argue that the framework {by Feldmann et al.}~\cite{highway} can be applied in both settings, yielding an analogous result as for bounded highway dimension.

        {%
        	{Finally}, we also show a limit of usefulness of relaxing the hop constraint by a constant amount, at least in the setting of non-metric distance functions. %
        	Extending a result of Manyem and Stallman~\cite{Manyem1996},
      we show that, unless $\Pol=\NP$, there is no chance of obtaining a $((1-c)\cdot\log n)$-approximation, for some constant $c>0$, with a Steiner tree that uses $k+\ell$ hops and comparing to an optimal solution that uses only $k$ hops.}

	\section{Preliminaries}
	\label{sec:Prelim}
	Let~$(V,\dis)$ be a metric induced by the graph~$G=(V,E)$. 
	{We assume w.l.o.g.\ that the metric is given by~$G$. 
	If this is not the case, we construct~$G$ as the complete graph on~$V$ where every 
	edge~$\{u,v\}$ has weight equal to the distance from $u$ to $v$. 
	In order to break ties consistently, we assume that shortest paths in~$G$ are unique. This can be 
	achieved by adding some sufficiently small random noise to the weight of each edge of~$G$. 
	We also assume that~$G$ is the {\em minimal graph inducing}~$(V,\dis)$. That is,  
	no edge in~$E$ can be removed without changing the length of some shortest path.
	A \kST for this modified instance is also optimal for the original instance.
	Furthermore, the minimal graph~$G$ inducing~$(V,\dis)$ is unique.}
	
	Given a metric,
	we can decide in polynomial time if it is a path metric, a tree metric, or a
	metric {with} treewidth $\omega$ for some constant $\omega\geq 1$. 
	To verify that $G$ is a path or a tree, we simply run a depth-first search. Moreover, for constant $\omega$, it can be decided in polynomial time whether~$G$ has treewidth $\omega$ by computing a treewidth decomposition~\cite{Bodlaender96}.

	We give two alternative representations of Steiner trees that are useful when working with partial solutions.
	Let~\stree be a Steiner tree on~$(V,\dis)$ with terminals $\Term\subseteq V$ and root~$r\in\Term$. Let $V_{\stree}\subseteq V$ with $\Term\subseteq V_{\stree}$ be the set of vertices in $\stree$.
	The tree $\stree$ can be viewed as a function
	mapping a vertex of~$V_{\stree} \setminus\{r\}$ to its immediate predecessor, i.e., its parent in~$\stree$. 
	More generally,
	for $U \subseteq V$, call a function $\anch : U\setminus\{r\} \longrightarrow
	V$ an \emph{anchoring on~$U$}. The \emph{anchor} $\anch(v)$ of vertex~$v$ represents its
	parent in \stree, and we set $\anch(w) = w$ if $w \notin V_{\stree}$.
	
	If \stree is of minimum cost, this additionally allows for the following representation. 
	Consider a function assigning to each vertex $v\in V_{\stree}$ its depth, i.e. the number of edges on the $r$-$v$ path in $\stree$. 
	Since a vertex $v\in V_{\stree} \setminus\{r\}$ of depth $x$ is anchored to the (uniquely determined) vertex of depth $x-1$ that is of minimum distance to $v$ w.r.t.\ $\dis$, this yields a complete representation of $\stree$.
	Generalizing again to subsets $U \subseteq V$, we call a function $\lab : U \longrightarrow \{0,1,\ldots, k\}\cup\{\emptylabel\}$ a \emph{labeling on $U$}.
	We call $\lab(w)$ the \emph{label of $w$} and set $\lab(w) = \infty$ if $w \notin V_{\stree}$.
	Note that this representation automatically enforces the $k$-hop condition.
	See Figure~\ref{fig:lap} for an example of a~\kST with the corresponding anchoring and labeling.

\begin{figure}[tb]
  \centering
\begin{tikzpicture}[xscale = .59, yscale = .73]
	\foreach \a/\b/\n/\p/\style in {9/5/r/r/terminal, %
		7/5/va/r/vertex, 7.5/2/vb/r/terminal, 10.4/3.3/vc/r/terminal, %
		4/4/va1/va/terminal, %
		2/3/va2/va1/vertex,
		0/1/va31/va2/terminal, 3/0/va32/va2/terminal,
		6/2.5/vb11/vb/vertex, 7/0/vb12/vb/vertex, %
		12/2/vc1/vc/terminal, %
		14/3/vc21/vc1/terminal, 14/0.8/vc22/vc1/vertex, 
		16.5/1/vc3/vc22/terminal}{
		\node[\style](\n) at (\a,\b) {};
	}
	\foreach \a/\b/\n/\p/\style in {9/5/r/r/terminal, %
		7/5/va/r/vertex, 7/2/vb/r/vertex, 10.4/3.3/vc/r/vertex, %
		4/4/va1/va/terminal, %
		2/3/va2/va1/vertex,
		0/1/va31/va2/terminal, 3/0/va32/va2/terminal,
		6/3/vb11/vb/vertex, 7/0/vb12/vb/vertex, %
		12/2/vc1/vc/vertex, %
		14/3/vc21/vc1/terminal, 14/0.8/vc22/vc1/vertex, 
		16.5/1/vc3/vc22/terminal}{	
		\draw[black, thick](\n) -- (\p);
	}

	{
		\draw[Steiner, utgreen](r) edge[bend right = 40] (va2);
		\draw[Steiner, utgreen](vc) edge[bend right = 20] (r);
		\draw[Steiner, utgreen](va1) edge[bend right = 20] (va2);
		\draw[Steiner, utgreen](va31) edge[bend left = 20] (va2);
		\draw[Steiner, utgreen](va32) edge[bend left = 10] (va2);
		\draw[Steiner, utgreen](vc) edge[bend left = 25] (vc1);
		\draw[Steiner, utgreen](r) edge[bend left = 25] (vb);
		\draw[Steiner, utgreen](vc21) edge[bend right = 15] (vc1);
		\draw[Steiner, utgreen](vc3) edge[bend right = 15] (vc1);
		\node[root]() at (vc) {};
	}
	\foreach \a/\b/\l in {9.2/5.4/$u_1$, 
10.9/3.6/$r$, %
8.2/1.8/$u_5$, %
		4.2/4.4/$u_3$, %
		1.8/3.4/$u$,
		-0.2/1.4/$u_4$, 3.2/0.4/$u_2$,
		11.6/1.8/$w$, %
		14.4/3.3/$w_1$,
		16.8/1.4/$w_2$}{	
		\node[]() at (\a,\b) {\l};
	}
	\node[anchor=south west]() at (18,-0.4) { \small
		$\begin{array}{ c | c | c }
		v & \anch(v) & \lab(v) \\
		\hline
		r & - & 0 \\
		u_1 & r & 1 \\
		u & u_1 & 2 \\
		u_2 & u & 3 \\
		u_3 & u & 3 \\
		u_4 & u & 3 \\
		u_5 & u_1 & 2 \\
		w & r & 1 \\
		w_1 & w & 2 \\
		w_2 & w & 2 \\
		w_3 & w & 2 \\
		x & x & \infty \\
		\end{array}$};
	\end{tikzpicture}
	\caption{A $3$-hop M\smash{\v S}T (thick, bent) %
		 with root $r$ and terminals (filled) $u_1,u_2,u_3, u_4,z,r,w,w_1,w_2$ on a tree metric (thin, straight) %
		 with unit-weight edges.
		Its cost is $12$.
		The table on the right describes the corresponding labeling \lab and anchoring \anch, where $x$ symbolizes vertices not used by the M\smash{\v S}T.}
	\label{fig:lap}
	\end{figure}                                                  	
 
	When $\stree$ is of minimum cost and $U=V$, we can easily compute an anchoring from a labeling or vice versa. 
	However, when considering partial solutions, i.e., when $U \varsubsetneq V$, this may not be possible.
	Thus, to retain the essential structural information, we utilize both representations simultaneously in this case.
	This motivates the following definition.
	
	\begin{definition}\label{dfn:lap}
		A pair $(\lab,\anch)$ is called a \emph{labeling-anchoring pair (\lap) on $U$} if the labeling \lab and anchoring \anch are consistent,
		i.e. for every {$u\in U\setminus\{r\}$} for 
		which $\anch(u)\in U$ and $\lab(u)\neq \emptylabel$,
		we have $\lab(u) = \lab(\anch(u))+1$. Moreover, if $\lab(u)
		= \emptylabel$ then $u\notin\Term$ and
		$\alpha^{-1}(u)=\{u\}$. 
	\end{definition}
	The \emph{cost} of a \lap $(\lab, \anch)$ is given by $\sum_{u\in U\setminus\{r\}}\dist{u,\anch(u)}$.
	In this sum, the term $\dist{u,\anch(u)}$ is called the \emph{cost to anchor} $u$.
	When $U \subsetneq V$, we may say \emph{partial} \lap to emphasize that the \lap only represents a portion of $\stree$, namely the edges between $U$ and its anchors.
	
	The representation as \lap is used to avoid the ambiguity that arises
	from simultaneously considering a Steiner tree $\stree$ and the tree-like graph $G$ that induces the underlying metric space. For example, in \Cref{sec:DPtrees}, both $\stree$ and $G$ are trees. 
	Throughout the paper, we represent Steiner trees as \laps. Hence, we use the term {\it anchor} to refer to a predecessor in~$\stree$ instead of {\it parent}.
	Moreover, when talking about distances or closeness, this always refers to distances in $G$. Given a point $v$ and a set $U\subseteq V$, denote by $\clo_v(U)$ the (unique) element of $U$ with minimum distance to $v$. For simplicity, we write $\clo_v(u,w)$ instead of $\clo_v(\{u,w\})$.

	In \Cref{sec:DPtrees,sec:DPboundedTW}, when querying a DP
	cell, a vertex with a desired label may not exist.
	To make these queries technically simple,
	we extend the vertex set of the metric to contain
	an auxiliary vertex, denoted by $\emptynode$. It is defined
	to have distance $\infty$ to all other vertices.
	In order to avoid the use of $k$ auxiliary vertices (one per label), we slightly {overload the} notation and assume that the equality $\lab(\emptynode)=i$ is correct for all $i\in[k]$, where $[k]\coloneqq\{1,2,\ldots, k\}$. Note that anchoring~$\emptynode$ incurs an infinite cost, so it will never be used in a $k$-hop Steiner tree.

	\section{The \texorpdfstring{\kST}{k-hop M\v{S}T} Problem in Path Metrics}
	\label{sec:DPline}

	Our first result is an efficient algorithm for \kST on path metrics:

	\begin{theorem}\label{thm:line}
		On path metrics, \kST can be solved exactly in time $O(kn^5)$.
	\end{theorem}

        We view a path metric as a set of vertices $V= \{v_1,v_2,\dots,v_n\}$ placed on the real line
        from left to right {by increasing index}, such that edges in the path correspond to consecutive vertices.

        On path metrics, we observe that there is no algorithmic difference between a \kST and a \kMST, since
        there exists a (uniquely defined) minimum-cost \kST $\topt=(\lab,\anch)$ rooted at $r\in V$ that only uses terminals.
        Indeed, if~$\topt$ contains a non-terminal vertex $v$, we may simply replace it by the next vertex on the line in the direction
	in which $v$ has the most edges (we break ties arbitrarily). This removes a non-terminal vertex without increasing the cost of $\topt$ or violating the $k$-hop condition.
	In this section, we therefore assume $\Term=V$.

	We give a recursive procedure
	which computes the \kMST, 
	and {we analyze its complexity} via dynamic programming.
	The goal is to first compute the internal (non-leaf) vertices of
	the \kMST and then add the cost of anchoring the
	leaves to the closest internal vertices. 

	\colorlet{stcolor}{green!40!black}
	\begin{figure}[tbhp]%
		\centering
		\begin{tikzpicture}[y=10pt, x=40pt]
		\draw[thick] (.5,0) -- (3.5,0);
		\draw[thick, dashed] (0,0) -- (4,0);
		\foreach \i in {1,2,3}
		\node[vertex, scale = .8] () at (\i, 0) {};

		{\color{stcolor}
			\node[vertex, fill=stcolor, label=left: $i$] (i) at (1, 2) {};
			\node[vertex,fill=stcolor, label=right: $s$] (h) at (2, 2.5) {};
			\node[vertex,fill=stcolor, label=right: $j$] (j) at (3, 1) {};
			
			\draw (h) -- (i);
			\draw [dotted] (i) -- (j);
		}
		
		\node[anchor=west] at (-1.5,0) {$G$};
		
		\node[anchor=west,stcolor] at (-1.5,2.2) {Steiner tree};
		\end{tikzpicture}
		\caption{On path metrics, the optimal \kST never anchors $j$ to $i$ if $\lab(i)>\lab(s)$ and $i<s<j$.}
		\label{fig:lineobservation}
	\end{figure} 	
	A key observation is the following. Fix an
	internal vertex $s$ {with} {label} %
	$\lab(s) < k$. It partitions the remaining vertex set into the vertices on the left of $s$ and those on the right of $s$. If a vertex $i$ to the left of $s$ {has} {label} %
	$\lab(i) > \lab(s)$, then in $\topt$, the vertex $i$ is never
	adjacent to a vertex to
	the right of $s$, see \Cref{fig:lineobservation}. This follows from
	the fact that such a vertex could be attached to~$s$ directly,
	decreasing the overall cost of $\topt$ without using more hops.
	
	We define a recursive expression $A[p,s,a,b]$ for $p\in \Nat$ and
	$s,a,b \in [n]$. It yields the minimum cost $p$-hop spanning tree $\stree$ rooted
	at $v_s$ that contains all vertices $v_i$ with~$i\in[a,b]$ and satisfies $s \notin [a,b]$. If $a>b$, let $[a,b]=\emptyset$.

	For $p\in \Nat$ and $s,a,b \in [n]$, define $A[p,s,a,b]$ as follows.
	\begin{enumerate}
		\item If $a > b$, then $A[p,s,a,b] = 0$.
		\item If $a = b$, then $A[p,s,a,a] = \dist{v_s,v_a}$.
		\item If $p = 1$, then $A[1,s,a,b] = \smash{\sum_{x \in [a,b]}} \dist{v_s,v_x}$ (all vertices anchored to $v_s$).
		\item If $p>1$, consider the right-most child $v_{s'}$ of $v_s$ in $\stree$ such that $s'\in [a,b]$. The sub-tree of~$\stree$ rooted at $v_{s'}$ covers all vertices $v_i$ with $i \in [c,b]$ for some $c \in [a,s']$. Thus, $A[p,s,a,b]$ is the sum of the cost of this subtree and that of all remaining subtrees of $v_{s}$ in $[a,c-1]$ plus the cost of connecting $v_s$ to $v_{s'}$. That is, $A[p,s,a,b]$ is defined as
	$$
	\small\min_{s' \in [a,b], \, c \in [a,s'-1]}\dist{v_{s'},v_s} + A[p, s, a, c-1] +
	A[p-1,s',c,s'-1] + A[p-1,s',s'+1, b]\,.
	$$
	\end{enumerate}
	See \Cref{fig:linerecursion}  for an illustration %
	where $b<s$. Note that in the last case, any recursive call can refer to an empty interval and incur zero cost.

	\begin{figure}[tb]\centering
		\colorlet{stcolor}{green!40!black}
		\begin{tikzpicture}[yscale = .75, xscale = .9, decoration={brace, amplitude=5pt}, pt/.style={minimum size = 25pt}]
		\def\sc{.5}
		\def\h{.35}
		\draw[thick, black!40] (1,0) -- (11,0);

		\node[vertex,scale = \sc] (jr) at (4.4, 0) {};
		\node[vertex,scale = \sc] (kr) at (7.1, 0) {};
		\node[vertex,scale = \sc] (ll) at (8.4, 0) {};
		\node[vertex] (y) at (1, 0)  {}; %
		\node[vertex] (b) at (10, 0) {}; %
		\node[pt] (lc) at (5,\h) {\small $c$};
		
		\draw[thick] (y) -- (jr);
		\draw[thick] (lc |- y) -- (kr);
		\draw[thick] (b |- y) -- (ll |- y);
		
		\node[vertex,scale = \sc] (jr) at (4.4, 0) {};
		\node[vertex,scale = \sc] (kr) at (7.1, 0) {};
		\node[vertex,scale = \sc] (ll) at (8.4, 0) {};
		\node[vertex] (y) at (1, 0)  {}; %
		\node[vertex] (b) at (10, 0) {}; %
		\node[pt] (lc) at (5,\h) {\small $c$};
		
		\node[pt] (y1) at (1,.7) {};
		\node[pt] (y2) at (1,-.1) {};
		\node[pt] (la) at (1,\h) {\small $a$};

		\node[vertex] () at  (5, 0) {}; %
		
		\node[vertex] () at (8, 0.0)  {}; %
		\node[pt] (lr1) at (8,\h+.05) {\small $s'$};
		
		\node[pt] (lb) at (10,.38) {\small $b$};
		
		\node[vertex] () at (11, 0)  {}; %
		\node[pt] (lr) at (11,\h) {\small $s$};

		\node[] (lrec1) at (2.7,-.8) {\scriptsize $A[p,s,a,c\!-\!1]$};
		\node[] (lrec2) at (6.1,-.8) {\scriptsize $A[p\!-\!1,s',c,s'\!-\!1]$};
		\node[] (lrec3) at (9.3,-.8) {\scriptsize $A[p\!-\!1,s',s'\!+\!1, b]$};
		
		\node[vertex, stcolor] (i) at (2, 1.5) {};
		
		\node[vertex,scale = \sc] () at (i |- y) {};
		\node[vertex,scale = \sc] (ir) at (2.6, 0) {};
		\node[vertex,scale = \sc] (jl) at (3.4, 0) {};
		\node[vertex,scale = \sc]  at (1.3, 0) {};
		\node[vertex,scale = \sc]  at (1.65, 0) {};
		\node[vertex,scale = \sc]  at (2.3, 0) {};
		\node[vertex, stcolor] (j) at (4, 1.5) {};
		\node[vertex,scale = \sc] () at (j |- y) {};
		\node[vertex, stcolor] (k) at (6.2, 1.1) {};
		\node[vertex,scale = \sc]  at (5.4, 0) {};
		\node[vertex,scale = \sc]  at (5.8, 0) {};
		\node[vertex,scale = \sc] () at (k |- y) {};
		\node[vertex, stcolor] (l) at (8.9, 1.1) {};
		\node[vertex,scale = \sc]  at (9.5, 0) {};
		\node[vertex,scale = \sc]  at (10.3, 0) {};
		\node[vertex,scale = \sc]  at (10.6, 0) {};
		\node[vertex,scale = \sc] () at (l |- y) {};
		\node[vertex, stcolor] (r1) at (8,  1.5) {};
		\node[vertex, stcolor] (s) at (11, 2.2) {};
		
		\begin{scope}[/pgf/decoration/raise=3pt]                
		\draw [decorate,line width=1pt] (jr |- y2) -- (y |- y2);
		\draw [decorate,line width=1pt] (kr |- y2) -- (lc |- y2);
		\draw [decorate,line width=1pt] (b |- y2) -- (ll |- y2);
		\end{scope}

		{\color{stcolor}
			
			\begin{scope}
			\draw[line width= .6 pt] (i) to [bend left=5] (s);
			\draw[line width= .6 pt] (j) to [bend left=2] (s);
			\draw[line width= .6 pt] (r1) to [bend right=5] (s);
			\draw[line width= .6 pt] (r1) to [bend right=2] (k);
			\draw[line width= .6 pt] (r1) -- (l);
			
			\draw[line width= .3 pt, fill = stcolor!20] (i) -- (y |- y1) -- (ir |- y1) -- (i);
			\draw[line width= .3 pt, fill = stcolor!20] (j) -- (jl |- y1) -- (jr |- y1) -- (j);
			\draw[line width= .3 pt, fill = stcolor!20] (k) -- (lc |- y1) -- (kr |- y1) -- (k);
			\draw[line width= .3 pt, fill = stcolor!20] (l) -- (ll |- y1) -- (b |- y1) -- (l);
			
			\end{scope}
		}
		\node[vertex, stcolor] () at (8.9, 1.1) {};
		\node[vertex, stcolor] () at (6.2, 1.1) {};
		\node[vertex, stcolor] () at (2, 1.5) {};
		\node[vertex, stcolor] () at (4, 1.5) {};
		
		\node[anchor=west] at (0,0) {$G$};
		
		\node[anchor=west,stcolor] at (0,2.2) {Steiner tree};

		\end{tikzpicture}
		\caption{Computation of $A[p,s,a,b]$ with three recursive calls.}
		\label{fig:linerecursion}
	\end{figure} 	\begin{proof}[{Proof of}~\Cref{thm:line}]
		Due to the key observation above, $A[p,s,a,b]$ correctly computes the minimum cost of a $p$-hop
		spanning tree $\stree$ with root $v_s$ and vertices $v_i$ with $i\in [a,b]$: For~$s'$ and $c$ as in $\topt$, there are no edges in $\topt$ between
		$[a,c-1]$, $[c,s'-1]$ and $[s'+1, b]$. Also, the recursive procedure only
		queries intervals $[a,b]$ with $s \notin [a,b]$.
		The cost of $\topt$ is~$A[k,r,0,r-1] + A[k,r,r+1,n]$.
		
		We dynamically compute the values $A[p,s,a,b]$ by iterating in an increasing manner over~$p$ in an outer loop and  the set of intervals $[a,b]$ in an inner loop, with shorter intervals having precedence.
		This is feasible, as a call of
		$A[p,s,a,b]$ recursively only queries values~$A[p',s',a',b']$ with
		$p' < p$ or $(b'-a')^+ < (b-a)^+$. 
		Assuming that all previous values are precomputed, the value of a cell $A[p,s,a,b]$ can be computed in time $O(n^2)$. Since there are only
		$kn^3$ possible values of $(p,s,a,b)$ to be queried, the
		total running time is bounded by~$O(kn^5)$.
	\end{proof}

	\section{The \texorpdfstring{\kST}{k-hop M\v{S}T} Problem in Tree Metrics}
	\label{sec:DPtrees}

	In this section, we construct a dynamic program for the \kST problem on tree metrics, formally proving the following:

        \begin{theorem}\label{thm:trees}
		On tree metrics, \kST can be solved exactly in time $n^{O(k)}$. 
	\end{theorem}

	Consider an instance of \kST with root $r\in\Term$ and metric $(V,\dis)$ induced by a tree~$T=(V,E)$. Without loss of generality, we consider $T$ to be rooted at $r$.
	For $v\in V$, denote by $T[v]$ the set
	of vertices
	in the subtree of $T$ rooted at $v$.
	
	We start by giving a high-level overview of our approach for computing the minimum~cost~$k$-hop Steiner tree~$\topt=(\lab, \anch)$.
	We use a dynamic program with cells $\bar A[v, \rho, \phi]$ indexed by a node $v \in V$ and vectors $\rho$ and $\phi$ of $k$ vertices each.
	Intuitively, $\rho$ and~$\phi$ represent \emph{anchoring guarantees} that convey information about the structure of~$\topt$ in relation to~$v$ and serve as possible points to which $v$ is anchored in~$\anch$. 
	Specifically, for each possible label~$i$, there is an anchoring guarantee inside {($\phi_i$)} and one outside {($\rho_i$)} of $T[v]$ acting as candidates for anchoring $v$ in $\topt$.
	We show that a cell $\bar A[v,\rho,\phi]$ computes a partial labeling-anchoring pair (\lap, recall \Cref{dfn:lap}) on $T[v]$ that is of minimum cost and respects the given anchoring guarantees.
	The cells are filled up in a bottom-to-top manner, starting at the leaves of the underlying tree $T$.
	Doing this consistently, while filling in correct anchoring guarantees, finally yields $\topt$.

	{A key property of a \kST that we implicitly use in the dynamic program
	recursion is that all nodes with label $i+1$ of a subtree $T[v]$ that are not
	anchored to a node of $T[v]$ are anchored to the same node: the closest node of
	label $i$ outside $T[v]$. Locating this node is the motivation of $\rho_i$.
	More generally, the objective of the \emph{anchoring guarantees} $\phi$ and
	$\rho$ is to constrain some node labels in order to convey this information
	consistently to the children $v_j$ of $v$: the $\rho_i$ associated to a $v_j$ is
	either the $\rho_i$ or the $\phi_i$ of $v$.}

	\subsection{Anchoring guarantees}
	Fix a vertex~$v\!\in\!V\setminus \{r\}$. Formally, { \emph{anchoring guarantees} of $v$ are of the form~$\phi(v) =
	\big(\phi_1(v),\dots,\phi_{k-1}(v)\big)$
	and~$\rho(v) = \big(\rho_1(v),\dots,\rho_{k-1}(v)\big)$
	with $\phi_{i}(v) \in T[v]$ and~$\rho_{i}(v) \in V\setminus T[v]$ for all $i\in [k-1]$.} Additionally, we allow the~$\phi_i(v)$ and~$\rho_i(v)$ to take the value $\emptynode$ and let $\rho_0(v) = r$ and \mbox{$\phi_0(v)=\emptynode$}.
	{Call anchoring guarantees of a vertex $v$ \emph{correct}, if~$\phi_i(v)$ and~$\rho_i(v)$ are the closest vertices to~$v$ with label~$i$ in a \kST inside resp.\ outside of~$T[V]$, or equal $\emptynode$ if no such vertex exists. See~\Cref{fig:rhophi} for an example where this is the case.
	If $v$ is part of the Steiner tree, then for an anchoring guarantee to be correct, the following must hold.}
	\begin{observation}\label{obs:correctAnch}
		{Given a {non-root} vertex~$v$ in $\opt = (\lab,\anch)$ and {its} correct anchoring guarantees $\rho(v)$, $\phi(v)$, there exists a unique label $i_v\in[k]$ such that $\phi_{i_v}(v) =v$. Moreover, this implies~$\ell(v) = i_v$ and $\anch(v) = \clo_{v}(\rho_{i_v-1}(v),\phi_{i_v-1}(v))$.}
	\end{observation}

\begin{figure}
	\centering
	\begin{tikzpicture}[xscale = .59, yscale = .73]
	\foreach \a/\b/\n/\p/\style in {9/5/r/r/terminal, %
		7/5/va/r/vertex, 7.5/2/vb/r/terminal, 10.4/3.3/vc/r/terminal, %
		4/4/va1/va/terminal, %
		2/3/va2/va1/vertex,
		0/1/va31/va2/terminal, 3/0/va32/va2/terminal,
		6/2.5/vb11/vb/vertex, 7/0/vb12/vb/vertex, %
		12/2/vc1/vc/terminal, %
		14/3/vc21/vc1/terminal, 14/0.8/vc22/vc1/vertex, 
		16.5/1/vc3/vc22/terminal}{
		\node[\style](\n) at (\a,\b) {};
	}
	\foreach \a/\b/\n/\p/\style in {9/5/r/r/terminal, %
		7/5/va/r/vertex, 7/2/vb/r/vertex, 10.4/3.3/vc/r/vertex, %
		4/4/va1/va/terminal, %
		2/3/va2/va1/vertex,
		0/1/va31/va2/terminal, 3/0/va32/va2/terminal,
		6/3/vb11/vb/vertex, 7/0/vb12/vb/vertex, %
		12/2/vc1/vc/vertex, %
		14/3/vc21/vc1/terminal, 14/0.8/vc22/vc1/vertex, 
		16.5/1/vc3/vc22/terminal}{	
		\draw[black, thick](\n) -- (\p);
	}

	{
		\draw[Steiner, utgreen](r) edge[bend right = 40] (va2);
		\draw[Steiner, utgreen](vc) edge[bend right = 20] (r);
		\draw[Steiner, utgreen](va1) edge[bend right = 20] (va2);
		\draw[Steiner, utgreen](va31) edge[bend left = 20] (va2);
		\draw[Steiner, utgreen](va32) edge[bend left = 10] (va2);
		\draw[Steiner, utgreen](vc) edge[bend left = 25] (vc1);
		\draw[Steiner, utgreen](r) edge[bend left = 25] (vb);
		\draw[Steiner, utgreen](vc21) edge[bend right = 15] (vc1);
		\draw[Steiner, utgreen](vc3) edge[bend right = 15] (vc1);
		\node[root]() at (vc) {};
	}
	{
		
		\foreach \a/\b/\l in {			4.05/4.45/{$v$}, %
			4.2/3.45/{${{\color{utgreen}\phi_3(\!v\!)}}$}}{	
			\node[]() at (\a,\b) {\l};
		}
		\draw[black!70, very thick] plot [smooth cycle, tension=1] coordinates {(-1.4,1.2) (4.1,-.2) (4.8,5.2)};
		\node[]() at (4,2.2) {$T[v]$};}

	{
		
		\node[fill=white, opacity=.9]() at (0.48,3.4) {{$v_1=\color{utgreen}\phi_2(\!v\!)$\hspace*{-3pt}}};
	}
	{
		\node[]() at (11.8,3.8){$r={\color{utred}\rho_0(v)}$};
		\foreach \a/\b/\l in {
			8.5/1.8/${\color{utred}\rho_2(v)}$,
			10/5.2/${\color{utred}\rho_1(v)}$
}{	
			\node[]() at (\a,\b) {\l};
		}
	}
	\end{tikzpicture}\vspace{-12 pt}
	\caption{{Anchoring guarantees $\rho(v)$, $\phi(v)$ for vertex $v$. The subtree $T[v]$ with respect to the underlying metric (thin, straight) is encircled. In this example, the depicted anchoring guarantees are indeed the correct choices for {an optimal $3$-hop} Steiner tree (thick, bent). To satisfy the anchoring guarantees, $v$~must be anchored to either $\phi_2(v)$ or $\rho_2(v)$.}
	\label{fig:rhophi}}
\end{figure}     %
 {This observation is of crucial importance for choosing the relevant anchoring guarantees when working with partial solutions.
	{Indeed, it means that correct anchoring guarantees completely determine the label and the anchor of $v$, hence there is no freedom nor complexity left to account for the cost to anchor $v$ in the dynamic program process.}
    Specifically, we are interested in partial \laps on~$T[v]$.}
	Given a \lap, denote by $\lambda_i(v)$ the vertex in~$T[v]$ with label $i$ closest to $v$ (or
	$\emptynode$ if no such vertex exists). 
	\begin{definition}\label{def:trees_P}
	{We define the set} $\mc P(v,\rho(v),\phi(v))$ to be the (possibly empty) set of \laps on $T[v]$ respecting the anchoring guarantees. That is, its elements $(\lab,\anch)$ satisfy:
	\begin{enumerate}[label=(\roman*)]
		\item For all $i$, we have $\phi_i(v) =\lambda_i(v)$. In particular, 
		if $\phi_i(v) = \phi_{j}(v)$ and $i\neq j$, then it holds that $\phi_i(v) = \emptynode$.
		\label{prop:Sp1}
		\item A vertex $w \in T[v]$ with $\lab(w)\neq\emptylabel$ is anchored to a vertex of $T[v]$ with label $\lab(w)-1$ or to $\rho_{\lab(w)-1}(v)$. Recall that $\lab(w)=\emptylabel$ implies $\alpha(w)=w$ (and $w\notin \Term$).\label{prop:Sp2}
	\end{enumerate}
\end{definition}
	Intuitively, $\mc P(v,\rho(v),\phi(v))$ represents all {relevant} 
	ways to extend a partial \lap $(\lab',\anch')$
	on $V\setminus T[v]$ to $V$ while respecting the anchoring guarantees: {all $\phi_i(v)$ are consistent with $\lambda_i(v)$ and all anchors outside of $T[v]$ are some $\rho_i(v)$, respecting labels.}
	Note that vertices of $T[v]$ are anchored either to another vertex in $T[v]$ or to some $\rho_i(v)$.
	Therefore, if $\rho_i(v)$ is used, it should be the closest vertex to $v$  outside~of~$T[v]$ for which $\lab'(\rho_i(v)) = i$.
	Assume $(\lab',\anch')$ is extended with minimum cost and consider the subtree $T[v_j]$ of a child $v_j$ of $v$.
	Its vertices are anchored either to a vertex of $T[v_j]$, or to a $\phi_i(v)$ (which may be in the subtree of a different child), or to
	a $\rho_i(v)$. The anchoring guarantees $\phi_i(v)$ are then necessary to
	determine the anchoring guarantees $\rho_i(v_j)$ for the children~of~$v$. Note that when defining $\mc P$, we do not require any constraint on the values of $\rho$.
	
	\subsection{The dynamic program}
	For $v\neq r$, denote by $A[v,\rho(v),\phi(v)]$ the minimum cost of a \lap on~$T[v]$ in $\mc P(v,\rho(v),\phi(v))$, or $\infty$ if none exists.
	{Assuming that these values for the children $v_1,v_2,\dots,v_p$ of $r$ and all possible anchoring guarantees {are known}, {we {observe} that the {cost} of {\opt} can be computed by evaluating} %
	\begin{align}\label{eq:tree-root-evaluation}
	{\sum_{j=1}^p} ~~\min_{\phi_i(v_j)\in T[v_j],~ \forall i\in[k-1]} ~~ A[v_j, \rho_\emptyset, \phi(v_j)],
	\end{align}
	where $\rho_\emptyset \coloneqq \{r,\emptynode,\dots,\emptynode\}$. This
	expression combines the minimum-cost LAPs on all subtrees of children of $r$,
	which, in an optimal solution, only have $r$ as an outer anchor. {It is
	crucial for the algorithm complexity to note that these partial LAPs can be
	optimized independently as there is no need to have an edge going through $r$
	in a Steiner tree solution: $r$ would be a better anchor as it is closer and
	less deep.} }
	
      {We now describe a dynamic program that computes {the values $A[v, \rho(v), \phi(v)]$.}
        Fix some vertex $v\neq r$ and d}enote by $v_1,v_2,\dots,v_p$ the children of $v$ in $T$.
	{Keeping \cref{obs:correctAnch} in mind,} we fill the cells $\bar A[v,\rho(v),\phi(v)]$ of our dynamic programming table according to the following recursive relation. For a vertex $v$ that is a leaf of $T$, we define
	\begin{align}
	\bar{A}[v,\rho(v),\phi(v)] &\coloneqq 
	\begin{cases}
	0\text{,}      & \text{if $v\notin \Term$ and $\phi_i(v)=v_\emptyset$ for all $i$;}\\
	\dist{v,\rho_{i_v-1}(v)}\text{,} & \text{if $\exists$ unique $i_v$, s.t.\ $\phi_{i_v}(v)=v$
		;}\\
	\infty\text{,} & \text{otherwise.}
	\end{cases}\label{eq:recursion2} 
	\intertext{ {The three cases correspond respectively to: a non-terminal node with no need to get a particular label; a node required by anchoring guarantees to get a label; inconsistent anchoring guarantees.}
		For non-leaf vertices, we define
	}
	\bar A[v,\rho(v),\phi(v)] &\coloneqq c_v + \smash{\sum_{j=1}^{p}} ~\min_{\phi_i(v_j)\in\bm\Phi_i(v_j), \forall i}~   \bar A[v_j,\bm\rho(v_j),\phi(v_j)]\,.\label{eq:recursion}
	\end{align}
	Here, $c_v$ denotes the cost of anchoring $v$ while $\bm\Phi_i(v_j)$ and
	$\bm\rho(v_j)$ encode which of the $n^{2k-2}$ possible anchoring
	guarantees of $v_j$ are consistent with that of $v$. The cells
	of each child are queried independently, {which is crucial for the algorithm complexity: the choice of some $\phi(v_j)$ does not impact the choice for other children of $v$}. Precise definitions of $\bm\Phi_i(v_j)$, $\bm\rho(v_j)$~and~$c_v$~follow.

	Let $\bm\Phi_i(v_j)$ be the subset of $T[v_j]$ consisting of all feasible choices for $\phi_i(v_j)$.
	Specifically, if $\phi_i(v)\in T[v_j]$, then $\bm\Phi_i(v_j)=\{\phi_i(v)\}$. Indeed, as the shortest $v$-$\phi_i(v)$ path passes through~$v_j$, node $\phi_i(v)$ must be the closest vertex to $v_j$ in~$T[v_j]$ with (already guaranteed)
	label~$i$.
	If~$\phi_i(v) = \emptynode$, we must have $\bm\Phi_i(v_j)=\{\emptynode\}$ or contradict Property \ref{prop:Sp1}.
	Otherwise, if~$\emptynode \neq \phi_i(v)\notin T[v_j]$, then $\bm\Phi_i(v_j)$ contains all~$w\in T[v_j]$ with~$\dist{v,w}\geq\dist{v,\phi_i(v)}$ and the auxiliary vertex $\emptynode$.
	A distance $\dist{v,w}<\dist{v,\phi_i(v)}$ would contradict the choice of $\phi_i(v)$ as the vertex in $T[v]$ with label~$i$ closest to~$v$.
	
	As for $\bm\rho_i(v_j)$, we define it to be the feasible choice for $\rho_i(v_j)$, which is (uniquely) determined as follows.
	If $\phi_i(v)\in T[v_j]$, then $\bm\rho_i(v_j)=\rho_i(v)$ since the shortest $v_j$-$\rho_i(v_j)$ path passes through $v$.
	Otherwise, we have $\bm\rho_i(v_j) = \clo_{v}(\rho_i(v),\phi_i(v))$.
	
	We now define $c_v$.
	If $v\notin \Term$ and no $\phi_i(v)$ equals $v$, then $c_v\coloneqq0$.
	Next, if there exists a unique~$i_v$ such that $\phi_{i_v}(v) = v$, let 
	$c_v \coloneqq \dist{v, \clo_v(\rho_{i_v-1}(v),\phi_{i_v-1}(v))}$. 
	In all other cases set $c_v \coloneqq \infty$, as the values of $\phi(v)$ are contradictory.
	
	\subsection{Analysis of the dynamic program} 
	The complexity {to evaluate Equations~\eqref{eq:tree-root-evaluation} and \eqref{eq:recursion}} is linear in the size of the table, i.e., $n^{O(k)}$, {and the complexity of \Cref{eq:recursion2} is $O(k)$.}
	Thus, in order to prove Theorem~\ref{thm:trees}, it remains to show the correctness of the dynamic program.
	{In this analysis, Properties \ref{prop:Sp1} and \ref{prop:Sp2} refer to \Cref{def:trees_P}.}
	
	\begin{proof}[Proof of~\Cref{thm:trees}] 
	By mathematical induction, we prove	that $\bar A[v,\rho(v),\phi(v)]$, 
	as defined in Equations~\eqref{eq:recursion2} and~\eqref{eq:recursion},
	is equal to $A[v, \rho(v), \phi(v)]$, for any node~$v\neq r$ and all possible anchoring guarantees~$\rho(v)$ {and} $\phi(v)$. 
	
	For the base step, i.e.~when $v$ is a leaf of $T$, we consider the three cases of
	Equation~\eqref{eq:recursion2}.
	If $v\notin\Term$ and $\phi_i(v)=v_\emptyset$ for all $i$, then clearly Properties~\ref{prop:Sp1} and \ref{prop:Sp2} are satisfied for the \lap that excludes $v$ from the Steiner tree, so $\bar A[v,\rho(v),\phi(v)] = A[v,\rho(v),\phi(v)] = 0$. Otherwise, there is at most one \lap that satisfies \ref{prop:Sp1} and \ref{prop:Sp2},
	namely the one that anchors~$v$ to~$\rho_{i_v-1}(v)$ if $i_v$ is defined. %
	It incurs a cost of $\dist{v,\rho_{i_v-1}(v)}$, as desired.
	If no such \lap exists, $\mc P(v,\rho(v),\phi(v))=\emptyset$ and both $A[v, \rho(v), \phi(v)]$ and $\bar A[v, \rho(v), \phi(v)]$ are infinite.
	This concludes the base step.

	Our induction hypothesis is that 
	\begin{equation*}
	\bar A[v',\rho(v'),\phi(v')] = A[v',\rho(v'),\phi(v')], \quad\quad \text{for all $v'\ne r$, $\rho(v')$, and $\phi(v')$}\,. \tag{IH}\label{eq:IH}
	\end{equation*}
	
	Now, for some non-leaf $v$, we assume that \eqref{eq:IH} holds for all descendants $v'\in T[v]\setminus\{v\}$ of~$v$ and prove that \eqref{eq:IH} holds for $v$ as well. For $v$, the recursive equation \eqref{eq:recursion} becomes
	\[
	\bar A[v,\rho(v),\phi(v)] \coloneqq c_v + \smash{\sum_{j=1}^{p}} ~\min_{\phi_i(v_j)\in\bm\Phi_i(v_j), \forall i}   A[v_j,\bm\rho(v_j),\phi(v_j)]\,. 
	\]

	If $c_v = \infty$, 
	then $\mc
	P(v,\rho(v),\phi(v))=\emptyset$, so both $A[v, \rho(v), \phi(v)] = \infty = \bar A[v, \rho(v), \phi(v)]$. From now on, assume that $c_v$ is finite.
        We prove \eqref{eq:IH} for $v$ by showing the two inequalities $A[v,\rho(v),\phi(v)] \geq \bar A[v,\rho(v),\phi(v)]$ and $A[v,\rho(v),\phi(v)] \leq \bar A[v,\rho(v),\phi(v)]$. {Once these two inequalities are proven, the proof of \Cref{thm:trees} is complete.}

	\begin{claim}
		{We have $A[v,\rho(v),\phi(v)] \geq \bar A[v,\rho(v),\phi(v)]$.}
	\end{claim}
	Consider the \lap $(\lab,\anch)$ which yields the value
	$A[v,\rho(v),\phi(v)]$. In particular, Properties~\ref{prop:Sp1} and \ref{prop:Sp2} are satisfied. If no such \lap exists, then $A[v,\rho(v),\phi(v)] = \infty$ and the inequality holds.
	For each child $v_j$ of $v$, set $\phi_i(v_j) = \lambda_i(v_j)$, which respects
	$\phi_i(v_j)\in \bm\Phi_i(v_j)$. Also, set $\rho(v_j) = \bm\rho(v_j)$ as defined above. We show that for each $v_j$, the restriction of the \lap $(\lab,\anch)$ to $T[v_j]$ belongs to $\mc{P}(v_j,\rho(v_j),\phi(v_j))$.
	
	Property~\ref{prop:Sp1} follows directly from the choice of $\phi_i(v_j)=\lambda_i(v_j)$.
	
	For Property~\ref{prop:Sp2}, consider a vertex $w\in T[v_j]$ which is not anchored to a vertex of~$T[v_j]$. We show that $\anch$ anchors $w$ to $\rho_{\lab_w}(v_j)$, with $\lab_w \coloneqq \lab(w)-1$. Note that by definition of~$\bm{\rho}(v_j)$, we have that $\rho_{\lab_w}(v_j)$ equals $\rho_{\lab_w}(v)$ or $\phi_{\lab_w}(v)$, so $\lab(\rho_{\lab_w}(v_j)) = \lab_w$. Since $\anch$ is an anchoring of minimal cost (with respect to the given anchoring guarantees), $w$ is anchored to the vertex
	$\anch(w) = \textstyle\clo_w\{x\in T[v]\cup \{\rho_{\lab_w}(v)\}\mid \lab(x)=\lab_w\}$, so $\anch(w) = \clo_{v_j}(\rho_{\lab_w}(v),\phi_{\lab_w}(v))$.
	If~$\phi_{\lab_w}(v)\in T[v_j]$, then $\bm\rho_{\lab_w}(v_j)= \rho_{\lab_w}(v) = \anch(w)$ as~$w$ is not anchored to a vertex in~$T[v_j]$.
	If~$\phi_{\lab_w}(v)\notin T[v_j]$, then $\bm\rho_{\lab_w}(v_j)= \clo_{v_j}(\rho_{\lab_w}(v),\phi_{\lab_w}(v)) = \anch(w)$, by definition of~$\bm \rho(v_j)$.

	Therefore, the \lap $(\lab,\anch)$ restricted to $T[v_j]$ belongs to $\mc{P}(v_j,\rho(v_j),\phi(v_j))$, so its cost is at least $A[v_j,\rho(v_j),\phi(v_j)]$.
	If $\lab(v)\neq\emptylabel$, then $\anch(v)=
	\clo_v(\rho_{i_v-1}(v),\phi_{i_v-1}(v))$ with cost~$c_v$, since the anchoring cost is minimized. 
	If $\lab(v)=\emptylabel$, then $c_v = 0$, so
	\begin{align*}
	A[v,\rho(v),\phi(v)] &= c_v + \smash{\sum_{j=1}^{p}} ~  A[v_j,\rho(v_j),\phi(v_j)] \geq \bar A[v,\rho(v),\phi(v)]\,.
	\end{align*}
	\begin{claim}
		{We have $A[v,\rho(v),\phi(v)] \leq \bar A[v,\rho(v),\phi(v)]$.}
	\end{claim}
	We assume $\bar A[v,\rho(v),\phi(v)]$ to be finite, otherwise the inequality trivially holds. 
	Consider the \laps that correspond to the values $A[v_j,\rho(v_j),\phi(v_j)]$ for which the
	value $\bar A[v,\rho(v),\phi(v)]$ is attained.
	We extend these {\lap}s on the subtrees $T[v_j]$ to $(\lab, \anch)$ on $T[v]$ in the following way. 
	{If $v\notin\Term$ and no $\phi_i(v)$ equals $v$, we let $\lab(v)=\emptylabel$ and $\alpha(v)=v$.} Otherwise, as $c_v\neq\infty$ by our assumption at the start of the proof, there exists a unique $i_v$ such that $\phi_{i_v}(v)=v$. We then let $\lab(v) = i_v$  and anchor~$v$ to
	$\clo_v(\rho_{i_v-1}(v),\phi_{i_v-1}(v))$. We show that this yields an element of~$\mc
	P(v,\rho(v),\phi(v))$.

	We first show Property \ref{prop:Sp1}. If $i_v$ is defined, $\phi_{i_v}(v) = v= \lambda_{i_v}(v)$ since $\lab(v) = i_v$. Consider~$\phi_i(v)$ for $i\neq i_v$. If $\phi_i(v)=\emptynode$, then all $\phi_i(v_j)=\emptynode${, too,} by definition of~$\bm\Phi_i(v_j)$. Thus, $\lambda_i(v_j) = \emptynode$ for all $j$ and $\lab(v)\neq i$, so $\lambda_i(v)=\emptynode = \phi_i(v)$. Otherwise, if~$\phi_i(v)\neq\emptynode$, there exists a $j_i$ with $\phi_i(v)\in T[v_{j_i}]$. Then, we have $\lambda_i(v_{j_i}) = \phi_i(v)$, and for all $j$, we have $\dist{v,\lambda_i(v_j)} = \dist{v,\phi_i(v_j)} \geq \dist{v,\phi_i(v)}$. Since $\lab(v)\neq i$, we obtain $\lambda_i(v) = \phi_i(v)$.
	
	It is easy to see that Property~\ref{prop:Sp2} holds as well. If we set $\anch(v)=v$, then $v\notin \Term$ and $\lab(v)=\emptylabel$. Otherwise, we define $\anch(v)$ to be either~$\rho_{i_v-1}(v)$ or~$\phi_{i_v-1}(v)\in T[v]$. Furthermore, any vertex $w$ of $T[v_j]$ is anchored either to a vertex in $T[v_j]\subseteq T[v]$ or to $\rho_{\lab(w)-1}(v_j)$, since the partial anchorings fulfill Property~\ref{prop:Sp2}. That means~$w$ is either anchored to a vertex of~$T[v]$ or, by definition of $\bm\rho(v_j)$, to $\rho_{\lab(w)-1}(v)$.
	
	In conclusion, $(\lab,\anch) \in\mc{P}(v,\rho(v),\phi(v))$, so its cost is at least $A[v,\rho(v),\phi(v)]$.
	\end{proof}

	\section{Metrics of Bounded Treewidth}
	\label{sec:DPboundedTW}

        We proceed to present our algorithm in full generality, building upon ideas from Section~\ref{sec:DPtrees} to obtain our main result {on metrics with bounded treewidth, which are defined as follows}. 

        {
        \begin{definition}\label{def:tw}
        	A graph $G=(V,E)$ is said to have \emph{treewidth} $\omega$, if there exists a tree $T_G = (B,E_B)$ whose nodes $b\in B$ are identified with subsets $S_b \subseteq V$, called \emph{bags}, satisfying: 
        	\begin{itemize}
        		\item[(i)] for each edge in $E$, there is a bag containing both endpoints, 
        		\item[(ii)] for each vertex in $V$, the bags containing it form a connected subtree of $T_G$, and 
        		\item[(iii)] each bag contains at most $\omega+1$ vertices.
        	\end{itemize}
       	The tree $T_G$ is called a \emph{tree decomposition} of $G$.
        \end{definition}
        }
\begin{figure}
	\centering
			\begin{tikzpicture}[xscale = .33, yscale = .4]

	\foreach \a/\b/\n/\p/\style in {9/5/r/r/terminal, %
		7/5/va/r/vertex, 7.5/2/vb/r/terminal, 10.4/3.3/vc/r/root, %
		4/4/va1/va/terminal, %
		2/3/va2/va1/vertex,
		0/1/va31/va2/terminal, 3/0/va32/va2/terminal,
		6/2.5/vb11/vb/vertex, 7/0/vb12/vb/vertex, %
		12/2/vc1/vc/terminal, %
		14/3/vc21/vc1/terminal, 14/0.8/vc22/vc1/vertex, 
		16.5/1/vc3/vc22/terminal}{
		\node[\style, opacity=.2](\n) at (\a,\b) {};
	}
	\foreach \a/\b/\n/\p/\style in {9/5/r/r/terminal, %
		7/5/va/r/vertex, 7/2/vb/r/vertex, 10.4/3.3/vc/r/vertex, %
		4/4/va1/va/terminal, %
		2/3/va2/va1/vertex,
		0/1/va31/va2/terminal, 3/0/va32/va2/terminal,
		6/3/vb11/vb/vertex, 7/0/vb12/vb/vertex, %
		12/2/vc1/vc/vertex, %
		14/3/vc21/vc1/terminal, 14/0.8/vc22/vc1/vertex, 
		16.5/1/vc3/vc22/terminal}{	
		\draw[black!10, thick](\n) -- (\p);
	}

	{
		\draw[Steiner, stcolor!10, thin](r) edge[bend right = 40] (va2);
		\draw[Steiner, stcolor!10, thin](vc) edge[bend right = 20] (r);
		\draw[Steiner, stcolor!10, thin](va1) edge[bend right = 20] (va2);
		\draw[Steiner, stcolor!10, thin](va31) edge[bend left = 20] (va2);
		\draw[Steiner, stcolor!10, thin](va32) edge[bend left = 10] (va2);
		\draw[Steiner, stcolor!10, thin](vc) edge[bend left = 25] (vc1);
		\draw[Steiner, stcolor!10, thin](r) edge[bend left = 25] (vb);
		\draw[Steiner, stcolor!10, thin](vc21) edge[bend right = 15] (vc1);
		\draw[Steiner, stcolor!10, thin](vc3) edge[bend right = 15] (vc1);
	}

				\node[vertex]() at (va) {};
				\node[terminal]() at (va1) {};
				\node[vertex]() at (va2) {};
				\draw[black, thick](va) -- (va1);
				\draw[black, thick](va2) -- (va1);

				\draw[black!70, very thick] plot [smooth cycle, tension=.8] coordinates {(7.7,6) (6.4,-0.7) (17.8,1)};
				\node[]() at (12.3,6) {$V\setminus\!T[v]$};
				
				\node[]() at (3.9,4.7) {$v$};
				\draw[black!70, very thick] plot [smooth cycle, tension=1] coordinates {(2.8,4.5) (4.6,3) (4.7,5.1)};
				\node[]() at (.2,4.2) {$T[v]\!-\!v$};
				
				\draw[black!70, very thick] plot [smooth cycle, tension=.9] coordinates {(-.6,1) (3.7,-.2) (2.3,3.7)};

			\begin{scope}[shift={({21},{0})}]
			\foreach \a/\b/\n/\p/\style in {9/5/r/r/, %
				7/5/va/r/vertex, 7.5/2/vb/r/vertex, 10.4/3.3/vc/r/vertex, %
				4/4/va1/va/terminal, %
				2/3/va2/va1/vertex,
				0/1/va31/va2/terminal, 3/0/va32/va2/terminal,
				6/2.5/vb11/vb/vertex, 7/0/vb12/vb/vertex, %
				12/2/vc1/vc/terminal, %
				14/3/vc21/vc1/terminal, 14/0.8/vc22/vc1/vertex, 
				16.5/1/vc3/vc22/terminal}{
				\node[\style, opacity=0](\n) at (\a,\b) {};
			}
				\draw[black!50, very thick] plot [smooth cycle, tension=.8] coordinates {(7.7,7) (6.4,-0.7) (18.8,2)};
				\node[]() at (13.5,7.2) {$V\setminus (S_b \cup C_b)$};
				
				\node[]() at (3.4,7) {$S_b$};
				\draw[black!50, very thick] plot [smooth cycle, tension=1] coordinates {(2,5.5) (4.28,3) (4.8,6.1)};
				
				\node[]() at (0,4.95) {$C_b$};
				\draw[black!50, very thick] plot [smooth cycle, tension=.9] coordinates {(-1,2) (3.7,-.2) (1.8,4.4)};
				
				\foreach \a/\b/\n in {7/5/a1, 7.2/6/a2, 6.3/4.1/a3, 6/3/a4,%
					4/4/b1, 3.75/4.8/b2, 3.2/5.6/b3,%
					2/3/c1, 1/4/c2, 2.7/2/c3%
				}{
					\node[vertex](\n) at (\a,\b) {};
				}
				\foreach \a/\b in {a1/b1, a2/b3, a3/b1, a3/b3, a1/b2, a4/b1,%
					b1/c1, b1/c3, b3/c2, b2/c2}{	
					\draw[black, thick](\a) -- (\b);
				}
				\foreach \a/\b/\n in {9.7/4.6/A1, 11/6/A2, 9.9/4/A3, 9/3/A4,%
					0/2.1/C1, 3.4/1/C2%
				}{
					\node[](\n) at (\a,\b) {};
				}
				\foreach \a/\b in {a1/r,a3/vc,a4/vb,a4/vb12,a2/A2,a1/A1,a3/A3,a3/A4,%
					c1/va31,c1/va32,c3/va32,c2/C1,c3/C2}{	
					\draw[black!15, dashed, thick](\a) -- (\b);
				}
			
			\end{scope}
			\end{tikzpicture}
	\caption{{Illustration of the key difference between tree metrics (left) and metrics of treewidth $\omega$ (right).
	While for tree metrics, we extend LAPs to subgraphs separated by only one cut-vertex, $v$, in the bounded treewidth case, cuts are formed by the bags $S_b$ of the tree decomposition and have size at most $\omega$.}}\label{fig:treeVsTw}
\end{figure}     We say that a metric $(V,\dis)$ \emph{has treewidth $\omega$}, if there exists a graph $G=(V,E)$ with treewidth $\omega$ that induces it.
	Our main result is as follows.

    \begin{theorem}\label{thm:treewidth}
    	{There exists an algorithm that solves the \kST problem exactly on metrics {with} treewidth $\omega$ and has a running time of $n^{O(\omega k)}$.}
	\end{theorem}
	{\cref{fig:treeVsTw} spotlights the main difference between \Cref{sec:DPtrees,sec:DPboundedTW}.
	In the former, we considered a subtree $T[v]$ of a tree metric $(V,\dis)$ and the question how to extend an LAP on $V\setminus T[v]$ to $T[v]$.
	Here, we could exploit the fact, that shortest paths between these sets pass through the root of the subtree, $v$, which forms a vertex cut of size $1$.
	In the bounded treewidth case, however, we consider the cuts formed by the bags $S_b$ of the tree decomposition. These contain up to $\omega$ vertices.
	While the approach conceptually is the same, the setting requires more care and is technically involved.}
	
	{Therefore, it will be useful to consider a more restricted variant of tree decompositions, see also~\cite{treewidth}, which we now define.}
	
	\begin{definition}
		{A \emph{nice tree decomposition} is a tree decomposition~$T_G = (B,E_B)$ in which}, w.r.t.\ a designated root $b_r\in B$, every node $b\in B$ {is of exactly} one of the following four types:
		\begin{itemize}	
		\item \emph{Leaf}: Its bag is empty, that is, $S_b=\emptyset$.

		\item \emph{Join node}: It has two children $b_1$ and $b_2$ with $S_b=S_{b_1}=S_{b_2}$.
		
		\item \emph{Forget node of $v$}: It has one child $b_1$ with $S_{b_1}=S_b\cup \{v\}$ and $v\notin S_b$.
		
		\item \emph{Introduce node of $v$}: It has one child $b_1$ with $S_{b_1}=S_b\setminus \{v\}$ and $v\in S_b$.	
        \end{itemize}
	\end{definition}

	\begin{figure}[tb]
		\newcommand{\convexpath}[2]{
			[   
			create hullnodes={#1}
			]
			($(hullnode1)!#2!-90:(hullnode0)$)
			\foreach [
			evaluate=\currentnode as \previousnode using \currentnode-1,
			evaluate=\currentnode as \nextnode using \currentnode+1
			] \currentnode in {1,...,\numberofnodes} {
				-- ($(hullnode\currentnode)!#2!-90:(hullnode\previousnode)$)
				let \p1 = ($(hullnode\currentnode)!#2!-90:(hullnode\previousnode) - (hullnode\currentnode)$),
				\n1 = {atan2(\y1,\x1)},
				\p2 = ($(hullnode\currentnode)!#2!90:(hullnode\nextnode) - (hullnode\currentnode)$),
				\n2 = {atan2(\y2,\x2)},
				\n{delta} = {-Mod(\n1-\n2,360)}
				in 
				{arc [start angle=\n1, delta angle=\n{delta}, radius=#2]}
			}
			-- cycle
		}
		\tikzset{create hullnodes/.code={
				\global\edef\namelist{#1}
				\foreach [count=\counter] \nodename in \namelist {
					\global\edef\numberofnodes{\counter}
					\node at (\nodename) [draw=none,name=hullnode\counter] {};
				}
				\node at (hullnode\numberofnodes) [name=hullnode0,draw=none] {};
				\pgfmathtruncatemacro\lastnumber{\numberofnodes+1}
				\node at (hullnode1) [name=hullnode\lastnumber,draw=none] {};
			},mynode/.style={outer sep=0pt,draw,shape=circle,minimum size=11mm,inner
				sep=0pt}}
		\tikzset{tw/.style={draw, ellipse, inner sep=3pt, minimum size=18pt, rectangle, rounded corners=10pt}}
		\tikzset{gr/.style={fill = white, draw, densely dashed, rectangle, rounded corners=3pt, inner sep=5pt, minimum size=5pt}}
		\tikzset{gredge/.style={draw, -, line width=6	pt, gray!60, shorten >=-5 pt, shorten <=-5 pt}}
		\colorlet{dkgreen}{green!50!black}
		\colorlet{dkred}{red!80!black}
		\begin{tikzpicture}[yscale = .9]
		\node[tw] at (0.8,-1.7) (c1) {$S_{b_1} = S_b$};
		\node[tw] at (-0.8,-1.7) (c2) {$S_{b_2} = S_b$};
		\draw[-latex] (0,-.7) edge (c1) (0,-.7) edge (c2);
		\node[tw, fill = white] at (0,-.7) (p) {$S_b$};
		\node[] at (0,0) {Join node};
		
		\node[gr] at (0,-3.2) (V) {$V\setminus (C_b\cup S_b)$};
		\node[gr] at (0,-4.5) (s) {$S_b=S_{b_1}=S_{b_2}$};
		\node[gr] at (-0.7,-6) (c1) {$C_{b_1}$};
		\node[gr] at (0.7,-6) (c2) {$C_{b_2}$};
		\begin{pgfonlayer}{bg}  
		\draw[gredge] (s) edge (V) edge (c1) edge (c2);
		\end{pgfonlayer}
		
		\node[align=center] at (-2.5, -1) {Bags in $T_G$};
		\node[align=center] at (-2.5, -4.5) {Connections\\ in $G$};v
		\draw[dkgreen,thick,densely dotted] \convexpath{c1.south west,c1.north west,c2.north east,c2.south east}{3pt};
		\node[dkgreen] at (1,-5.3) {$C_b$};
		
		\end{tikzpicture}\hfill
		\begin{tikzpicture}[yscale = .9]
		\node[tw] at (0,-.7) (p) {$S_b$};
		\node[tw] at (0,-1.7) (c) {$S_{b_1} = S_b\cup\{v\}$};
		\draw[-latex] (p) edge (c);
		\node[] at (0,0) {Forget node};
		
		\node[gr] at (0,-3.2) (V) {$V\setminus (C_b\cup S_b)$};
		\node[gr] at (-0.6,-4.5) (s) {$~~~S_{b}~~~$};
		\node[gr,solid] at (1,-5.5) (v) {$v\vphantom{S_b}$};
		\node[gr] at (-0.6,-6) (c1) {$C_{b_1}$};
		
		\draw[dkgreen,thick,densely dotted] \convexpath{c1.south east, c1.south west,c1.north west,v.north west,v.north east,v.south east}{3pt} node[pos=0.55, right, yshift=-10pt, xshift = 13 pt] {$C_b$};
		\begin{pgfonlayer}{bg}  
		\draw[gredge] (s) edge (V) edge (c1) edge (v) (v) edge (c1);
		\draw[gredge] (s) edge (V) edge (c1) edge (v) ;
		\end{pgfonlayer}
		\fill[dkred,thick,densely dotted,opacity=0.15] \convexpath{s.south west,s.north west,s.north east,v.north east,v.south east,v.south west}{2pt}; 
		\node[dkred] at (1, -4.5) {$S_{b_1}$};
		\end{tikzpicture}\hfill
		\begin{tikzpicture}[yscale = .9]
		\node[tw] at (0,-.7) (p) {$S_b$};
		\node[tw] at (0,-1.7) (c) {$S_{b_1} = S_b\setminus\{v\}$};
		\draw[-latex] (p) edge (c);
		\node[] at (0,0) {Introduce node};
		\node[gr] at (0,-3.2) (V) {$V\setminus (C_b\cup S_b)$};
		\node[gr] at (-0.6,-4.8) (s) {$~~~S_{b_1}~~~$};
		\node[gr,solid] at (1.2,-4.4) (v) {$v\vphantom{S_b}$};
		\node[gr] at (0,-6) (c1) {$C_{b_1} = C_b$};
		\begin{pgfonlayer}{bg}  
		\draw[gredge] (s) edge (V) edge (c1) edge (v) (v) edge (V);
		\end{pgfonlayer}
		\draw[dkgreen,thick,densely dotted,opacity=1] \convexpath{s.south east,s.south west,s.north west,v.north west,v.north east,v.south east}{2pt}; 
		\node[dkgreen] at (1.1,-5.25) {$S_b$};
		\phantom{
			\draw[dkgreen,thick,densely dotted] \convexpath{c1.south east, c1.south west,c1.north west,v.north west,v.north east,v.south east}{3pt} node[pos=0.5, below right] {$C_b$};
		}
		\end{tikzpicture}
		\caption{Types of bag nodes in a \emph{nice} tree decomposition and possible edges in $G$.}
		\label{fig:tw}
	\end{figure} 	{For an illustration, we refer to \Cref{fig:tw}.} By \Cref{def:tw}, Property~(ii), a vertex in $V$ may have several introduce nodes but at most one forget node.
	Let $C_b$ be the union of the bags $S_{b'}$ for all descendants 
	$b'$ of $b$, excluding vertices in $S_b$. Property~(ii) implies that there is no edge between $C_b$  and \mbox{$V\setminus(S_b\cup C_b)$,} %
	and that, for a join node, $C_{b_1}\cap C_{b_2}=\emptyset$.
	
	Given a graph with treewidth $\omega$, it is possible to compute a nice tree decomposition with $|B| = O(n\omega)$ in polynomial time~\cite{treewidth}. W.l.o.g.\ our input is a nice tree decomposition~$T_G$.

	Choose a root node~$b_r$ 
	whose bag contains the root~$r$ of the \kST which we aim to compute.
	To extend the dynamic programming approach from Section~\ref{sec:DPtrees} to nice tree decompositions, we again compute cells in a bottom-up fashion, now in~$T_G$.
	{As mentioned above, a} key difference
	lies in the fact that, here, a node $b$ in $T_G$ corresponds to several vertices in~$G$, so we require anchoring guarantees for every vertex in~$S_b$.
	{A DP cell, indexed by a bag~$b$ and $O({\omega k})$ anchoring guarantees, computes a minimum cost LAP on $C_b$ that respects these guarantees.}
	Thankfully, the structure of the nice tree decomposition enables us to recurse in an organized manner and construct the cells consistently. Join nodes combine previous results. Forget nodes decide the label and anchoring of the corresponding vertex and possibly new anchoring guarantees needed due to forgetting it. Introduce nodes deduce anchoring guarantees about the introduced vertex from previous knowledge. 

	{We again implicitly rely on structural properties of Steiner trees to make use of no more than $O(\omega k)$ anchoring guarantees describing the structure outside of $C_b$. Consider a \kST where two nodes $v_1$ and $v_2$ of $C_b$ with the same label $i+1$ are anchored to nodes outside of $C_b$, and the shortest paths from $v_1$ to $\alpha(v_1)$ and from $v_2$ to $\alpha(v_2)$ go through the same node $u\in S_b$. Then, we must have $\alpha(v_1)=\alpha(v_2)$. Locating this node is the objective of the anchoring guarantee $\rho_i(u)$, and the anchoring guarantees $\phi_i(u)$ allow us to convey this information consistently through the dynamic program recursion.}
	
	\subsection{{Anchoring Guarantees}}
	Fix a bag $b\in B$. Its \emph{anchoring guarantees} are {of the form} $\phi(b) = \{\phi_i^u(b)\mid  i\in[k-1] \wedge u\in S_b\}$  and $\rho(b) = \{\rho_i^u(b) \mid i\in[k-1]  \wedge u\in S_b\}$, with all $\phi_i^u(b)\in \{\emptynode\}\cup C_b$ and $\rho_i^u(b)\in  \{\emptynode\}\cup V\setminus C_b$. We additionally set  $\phi_0^u(b)=\emptynode$ and $\rho_0^u(b)=r$ for all $u$, see \Cref{fig:twrhophi}. We use these anchoring guarantees to define a subset of partial LAPs on $C_b$ analogously to \Cref{def:trees_P}. 
	{
	\begin{definition}
	We define the set $\mc P(b,\rho(b),\phi(b))$ to be the (possibly empty) set of partial LAPs on $C_b$ respecting the anchoring guarantees. That is, its elements $(\lab, \anch)$ satisfy:
	\begin{enumerate}[label=(\roman*$'$)]
		\item $\phi_i^u(b)$ is {the} closest vertex to $u$ in $C_b$ with label $i$ (or $\emptynode$ if no such vertex exists);\label{prop:Sp1'}	
		\item 
		Each vertex $u$ of $C_b$ with $\lab(u)\neq\emptylabel$ is anchored either to a vertex of $C_b$ with label $\lab(u)-1$ or to $\rho_{\lab(u)-1}^w(b)$ for some $w\in S_b$.
		\label{prop:Sp2'}
		\item 
		For all $i$ and $u, w\in S_b$, we have  $\dist{u,\rho_i^u(b)} \leq  \dist{u,\rho_i^w(b)}$.
		\label{prop:Sp3'}
	\end{enumerate}
	\end{definition}
	}
	Intuitively, $\mc P(b,\rho(b),\phi(b))$ represents all relevant ways to extend a partial LAP on $V\setminus C_b$ to $V$.
 Vertices of $C_b$ are anchored either to a vertex of $C_b$ or to some $\rho_i^u(b)$.  {An addition compared to \Cref{sec:DPtrees} is the last item, which requires that the $\rho_i^u(b)$ are consistent between each other. If the inequality was reversed, then $\rho_i^w(b)$ would have been a better choice for~$\rho_i^u(b)$. Indeed, the intuitive objective of $\rho_i^u(b)$ is to select the vertex of $V\setminus C_b$ with label~$i$ closest to $u$. As this is a guarantee too strong to be maintained in some cases where $\rho_i^u(b)$ is not useful as it is too far compared to $\phi_i^u(b)$, we only require this weaker guarantee of consistency among the other anchoring guarantees.}

	\begin{figure}[tb]\centering
		\begin{tikzpicture}[scale= 1,vt/.style ={draw,  cross out, minimum size=3pt, inner sep=1pt}, 
		every label/.style={fill=white, inner sep=1pt, label distance=3pt},
		subset/.style={set,rounded corners, dashed, thick},
		zig/.style={decorate, decoration={coil, aspect=0, segment length=15pt, amplitude=1pt}}]
		\colorlet{set}{green!50!black}
		\draw[subset] (-1, 0) rectangle (-5, 2);
		\node[set,anchor=north west] at (-5,2) {$V\setminus (S_b \cup C_b)$};
		\draw[subset] (0, 0) rectangle (2, 2); 
		\node[set,anchor=north west] at (0,2) {$S_b$};
		\draw[subset] (3, 0) rectangle (7, 2);
		\node[set,anchor=north east] at (7,2) {$C_b$};
		
		\node[vt, label=below:{$u=\rho^u_2=\rho^w_2$}] at (0.5,0.3) (u) {};
		\node[vt, label=above:{$w=\rho^w_3$}] at (1.5,1.4) (v) {};
		
		\node[vt, label=below:$\rho^u_3$] at (-1.4,0.3) (ru3) {};
		\node[vt, label=above:$\rho^w_1$] at (-1.7,1.4) (rv1) {};
		\node[vt, label=left:$\rho^u_1$] at (-3.7,1.3) (ru1) {};
		\node[vt, label=below:{$\rho^u_0=\rho^w_0=r$}] at (-4,0.7) (r) {};
		
		\node[vt, label=below:$\phi^u_1$] at (3.4,0.3) (pu1) {};
		\node[vt, label=right:{$\phi^u_2=\phi^w_2$}] at (5.5,0.6) (pu2) {};
		\node[vt, label=above:$\phi^w_1$] at (3.7,1.4) (pv1) {};
		
		\draw[zig, black!40, densely dotted, thick] (u.center) -- (v.center) --(rv1.south) -- (ru1.west) -- ($(r.center) + (0,0.1)$) -- (ru3.center) -- (u.center) -- (pu1.center) -- ($(pu2.center) + (-0.23,-0.01)$) -- (pv1.center) -- (v.center);
		
		\end{tikzpicture}
		\caption{Possible values of $\rho_i{(b)}$ and $\phi_i{(b)}$ {w.r.t.\ }two vertices $u$ and $w$. Note that $\phi_3^u =\phi_3^w=\emptynode$.}
		\label{fig:twrhophi}
	\end{figure} 

	\medskip

	\subsection{{The dynamic program}}\label{sec:treewidth_DP}
	{Let $A[b,\rho(b),\phi(b)]$ be the minimum cost of a partial LAP on $C_b$ in $\mc P(b,\rho(b),\phi(b))$, or $\infty$ if this set is empty. 
	Note that if Property~\ref{prop:Sp3'} is satisfied and $b$ is a leaf, then $A[b,\rho(b),\phi(b)]=0$.}

        {First, let us argue how knowledge of the values
          $A[b,\rho(b),\phi(b)]$ allows us to compute the cost of the optimal
          $k$-hop Steiner tree $\opt$.

          We have already fixed one bag $b_r$ of the nice tree
          decomposition containing $r$ to serve as the root of the
          decomposition.  We can imagine $\opt$ being computed in a two phase
          process: First, the optimal labeling and anchoring on vertices of the
          set $S_{b_r}$ is selected, with anchoring choices even outside $S_{b_r}$
          being taken into consideration. Next, we compute the optimal extension of
          this anchoring and labeling pre-selection on the rest of the vertices,
          namely $V \setminus S_{b_r}$. Such a process clearly yields the cost of $\opt$.

          As any \lap in $\mc P(b_r, \rho(b_r), \phi(b_r))$ by definition covers
          the entire $C_{b_r} = V \setminus S_{b_r}$ and $A[b_r, \rho(b_r), \phi(b_r)]$ is the
          minimum cost of the \lap{}s therein, we can repeat the above two-phase process formally:
      
          Assume we are given a partial labeling $\bar \lab$ on $S_{b_r}\setminus\{r\}$ and values $\phi(b_r)$.
          For all~$v\in S_{b_r}$, we define
          $\rho_i^{v}(b_r)$ as the vertex $w$ closest to $v$ for which
          $\bar \lab(w) = i$. The minimum cost of a \lap extending $\bar \lab$ that respects $\phi(b_r)$
          is equal to
          \begin{equation}\label{eq:treewidth-root-call}
            A[b_r,\rho(b_r),\phi(b_r)] + \sum_{v\in S_{b_r}\setminus\{r\}} \dist{v,\clo_{v}(\phi_{\bar \lab(v)-1}^v(b_r), \rho_{\bar \lab(v)-1}^v(b_r))}.
          \end{equation}
          As argued above, picking the $\bar \lab$ and $\phi$ that minimize this value gives us the cost of \opt.
        }
        
	In the following, we define in a recursive procedure how to fill the cells $\bar A[b,\rho(b),\phi(b)]$ of our dynamic programming table $\bar A$ for each node $b\in B$.
	The goal will be to again show that $\bar A= A$.
	First, as an easy special case,
	if Property~\ref{prop:Sp3'} is not respected by $\rho(b)$,
	we set $\bar A$ to infinity, which matches with the fact that $\mc P(b, \rho(b), \phi(b))$
	is empty and $A = \infty$.
	Next, we describe how to compute~$\bar A$
	depending on the type of the node $b$ when \ref{prop:Sp3'} is respected.\medskip
	
	\noindent
	\textit{Leaves:} Node $b$ has no child and $S_b=\emptyset$. We set $\bar{A}[b, \rho(b),\phi(b)] = 0$.
	\medskip
	
	\noindent
	\textit{Join nodes:} Node $b$ has children $b_1,b_2$ with
	$S_{b_1} = S_{b_2} = S_b$ and $C_{b_1}\cup C_{b_2} = C_b$ %
	and $C_{b_1}\cap C_{b_2} = \emptyset$. 
		Intuitively, the objective is to independently query partial solutions on each~$C_{b_j}$.
		We compute sets of possible values for $\rho^u_i(b_j)$ and $\phi^u_i(b_j)$ which define sets of partial LAPs on each $C_{b_j}$ respecting such guarantees. These possible values are determined such that the minimum cost of a combination of any two partial LAPs on $C_{b_1}$ and $C_{b_2}$ in these sets equals $A[b,\rho(b),\phi(b)]$.
	Here, the $\rho^u_i(b_j)$ need to be equal to the closest anchoring guarantee outside of $C_{b_j}$. The $\phi^u_i(b_j)$ may take any
	value not contradicting~$\phi(b)$. Specifically, for both $j\in\{1,2\}$, $i \in [k-1]$ and $u\in S_b$:
	\begin{itemize} 
		\item We set {$\bm\rho_i^u(b_j) =\clo_u\{\{\rho_i^u(b)\}\cup\{\phi_i^w(b)~|~ w\in S_b~\wedge~\phi_i^w(b)\notin C_{b_j}\}\}$}.
		\item If $\phi_i^u(b)\in C_{b_j}$, then we set $\bm\Phi_i^u(b_j) = \{\phi_i^u(b)\}$. 
		Otherwise, we set
		$$
		\smash{		\bm\Phi_i^u(b_j) = \{x\in C_{b_j}\cup\{\emptynode\}\mid \text{for all } z \in S_b, \text{ we have  }\dist{z,\phi_i^z(b)} \leq \dist{z,x} ~~~\text{($\star$)}
			\}}\,,
		$$
		where ($\star$)
		ensures that $\phi_i^z(b)$ is the vertex in $C_b$ that is closest to $z$.
	\end{itemize}
	We then define
	\begin{equation}
	\label{eq:twjoin}
	{\bar A[b,\rho(b),\phi(b)] = \sum_{j\in \{1,2\}} ~\min_{\phi_i^u(b_j)\in\bm\Phi_i^u(b_j),~ \forall i\in[k-1], \,u\in S_{b_j}}~   { \bar A[b_j,\bm\rho(b_j),\phi(b_j)]}}\,.
	\end{equation}
	
	\noindent
	\textit{Forget node of $v$:} We have $S_{b_1} = S_b\cup \{v\}$ and $C_{b_1} = C_b\setminus\{v\}$. There is no edge between $v$ and $V\setminus (C_b \cup S_b)$. 
	In this node, we want to define the label $i_v$ of $v$ and the corresponding anchoring of $v$.
	However, $\phi(b)$ may not contain sufficient information for deciding $i_v$, since~$v$ can be far away from $S_b$. We therefore need to consider all possible values for $i_v$, that is all values that are consistent with the guarantees $\phi(b)$.
	We first define the set $\bm I_v$ of possible labels of $v$ that do not contradict $\phi(b)$, then 
	proceed to define possible values of $\phi(b_1)$, $\rho(b_1),$ and finally, we express the cost to anchor $v$.
	
	Let $\bm I_v$ be the set of labels $i$
	such that for all $u\in S_b$, we have $\dist{u,v} \geq
	\dist{u,\phi_{i}^u(b)}$ and for all~$i'\neq i$, we have
	$\phi_{i'}^u(b)\neq v$. In other words, if there is a label $i$ and~$u\in S_b$ with~$\phi_{i}^u(b)=v$, then $\bm I_v$ cannot contain
	any other label: In order to respect the guarantee~$\phi_{i}^u(b)=v$, we must have $i_v=i$. Moreover, if there exists some $u\in S_b$ and $i$
	such that~$\phi_{i}^u(b)$ is further from~$u$ than $v$, then $\bm I_v$
	cannot contain $i$ as it would contradict the definition of~$\phi_i^u(b)$. If~$v\notin \Term$ and no $\phi_{i}^u(b)$ equals $v$,
	we include $\emptylabel$ in $\bm I_v$ as $v$ does not need to have a finite label in order to respect the guarantees $\phi_i(b)$.
	If $\bm I_v$ is empty,
	set $\bar A[b,\rho(b),\phi(b)]$ to be infinite since it is impossible to label $v$ while respecting the guarantees $\phi_i(b)$. 
	Assume now that $\bm I_v$ is not empty.

	The values $\phi_i^u(b_1)$ can take any value in {$C_{b_1}$} not contradicting $\phi(b)$.
	Specifically, for~$u\in S_b$, if $\phi_i^u(b)\neq v$ let $\bm\Phi_i^u(b_1)= \{\phi_i^u(b)\}$, and if $\phi_i^u(b)= v$, let $$\smash{\bm\Phi_i^u(b_1)= \{x\in {C_{b_1}\cup\{\emptynode\}}\mid  \dist{u,v} \leq \dist{u,x}\}}\,.$$
	Indeed, if $\phi_i^u(b)=v$, then we need to provide a new guarantee for $\phi_i^u(b_1)$, as $v\in S_{b_1}$, which must be further from $u$ than $v$.  
	We also define
	\begin{align*}
	\smash{\bm\Phi_i^v(b_1) = \{x\in {C_{b_1}}\cup\{\emptynode\}\mid
		\text{for all } u \in S_b,\text{ we have }\dist{u,\phi_i^u(b)} \leq \dist{u,x} ~~~\text{($\star$)}
		\}.}
	\end{align*}
	Again, ($\star$) must be satisfied since $\phi_i^u(b)$ is the vertex in $C_b$ which is closest to $u$.
	
	For the remainder, fix some $i_v \in \bm I_v$.
	In the case where $i_v=\emptylabel$, we need not consider~$\rho_{i_v}$'s.
	Otherwise, any path from $v$ to a vertex in $V\setminus C_b$ passes through $S_b$.
	Therefore, $\rho_i^v(b_1)$ is determined by $\bm\rho_i^v(b_1)=\clo_v\{\rho_i^u(b)\mid u\in S_b\}$ for $i\neq i_v$, and $\bm\rho_{i_v}^v(b_1)=v$.                                                                                                                                       
	Similarly, for~$u\in S_b$, let $\bm\rho_{i}^u(b_1) = \rho_{i}^u(b)$ for $i\neq i_v$, and $\bm\rho_{i_v}^u(b_1) = \clo_u\{\rho_{i_v}^u(b), v\}$.
                                                                         
	Additionally, we charge a cost of $c_{i_v}$ for anchoring $v$. If $i_v=\emptylabel$ then set $c_{i_v}\coloneqq 0$. Otherwise, set $c_{i_v} \coloneqq \dist{v, \clo_v\{\phi_{i_v-1}^v(b_1),\rho_{i_v-1}^v(b_1)\}}$.
	
	We then define, with $\bm\rho(b_1)$ depending on $i_v$ and $c_{i_v}$ depending on $\phi_i^u(b_1)$,
	\begin{equation}
	\label{eq:twforget}
	{\bar A[b,\rho(b),\phi(b)] = \min_{i_v\in \bm I_v} ~ \min_{\phi_i^u(b_1)\in\bm\Phi_i^u(b_1), ~\forall  i\in[k-1], \,u\in S_{b_j}}~ \Big(  c_{i_v} + {\bar A[b_1,\bm\rho(b_1),\phi(b_1)]\Big)}}\,.
	\end{equation}

	\smallskip
	\noindent
	\textit{Introduce node of $v$:} In this case, $b$ has one child $b_1$ with $S_{b_1} = S_b\setminus \{v\}$ and  $C_{b_1} = C_b$. There is no edge between $v$ and $C_b=C_{b_1}$
	as $v\notin S_{b_1}\cup\, C_{b_1}$, see \Cref{fig:tw}.
	If there is an $i$ with $\phi_i^v(b)\neq \clo_v\{\phi_i^u(b)\mid u\in S_{b_1}\}$ then $\bar A[b,\rho(b),\phi(b)]$ is infinite since the shortest $v$-$\phi_i^v(b)$ path has to pass through a vertex of $S_b$ by the above observation.
	Otherwise, the guarantees do not change, so we define $\rho(b_1) = \rho(b)$, $\phi(b_1)=\phi(b)$, and we set
	\begin{equation}
	\label{eq:twintro}
	\bar A[b,\rho(b),\phi(b)] = {\bar A[b_1,\rho(b_1),\phi(b_1)]}\,.
	\end{equation}

	\subsection{Analysis of the dynamic program}
	{Given the above definitions and the size of the dynamic programming table,
	one can check that~$\bar A$ {as well as \Cref{eq:treewidth-root-call}} can be computed in time~$n^{O(\omega k)}$.}
	It remains to show the correctness of the dynamic program.

	\begin{proof}[Proof of \cref{thm:treewidth}]
	By mathematical induction on the nice tree decomposition $T_G$, we prove
	that the value of the dynamic program cell $\bar A[b,\rho(b),\phi(b)]$ as defined in \Cref{sec:treewidth_DP}
	is equal to the minimum cost $A[b, \rho(b), \phi(b)]$ of a LAP in
	$\mc P(b, \rho(b), \phi(b))$.
	
	If $b$ is a leaf, then depending on Property~\ref{prop:Sp3'}, both values are either $\infty$ or zero and the result holds.	
	Otherwise, we assume by induction hypothesis that for all children $b_j$ of $b$ and all possible anchoring guarantees $\rho(b_j)$ and $\phi(b_j)$, we have $A[b_j,\rho_j(b),\phi_j(b)] = \bar A[b_j,\rho_j(b),\phi_j(b)]$. We now show this equality for node $b$, by proving both inequalities separately and according to the type of $b$, { concluding the proof of \cref{thm:treewidth}.}

	\begin{claim}\label{claim:TW_IEQ1}
		{We have	$A[b,\rho(b),\phi(b)] \geq \bar A[b,\rho(b),\phi(b)]$.}
	\end{claim}
	Consider a bag $b$ and any values $\rho(b)$ and $\phi(b)$. 
	Consider a \lap $(\lab,\anch)\in \mc P(b,\rho(b),\phi(b))$ for which the value
	$A[b,\rho(b),\phi(b)]$ is attained. If no such \lap exists, then $A[b,\rho(b),\phi(b)] = \infty$ and the inequality holds.
	There are three different cases depending on the type of $b$. For each
	case, we focus on a bag node $b_j$ child of $b$ and define values
	$\rho(b_j)$ and $\phi(b_j)$. We show that the restriction of
	$(\lab, \anch)$ to $C_{b_j}$ belongs to $\mc P(b_j, \rho(b_j), \phi(b_j))$. We then show that the value of cell $\bar A[b_j, \rho(b_j), \phi(b_j)]$ was considered in the computation of $\bar A[b, \rho(b), \phi(b)]$, i.e., each~$\phi_i^u(b_j)$ belongs to
	the corresponding $\bm\Phi_i^u(b_j)$ and $\rho(b_j) = \bm \rho(b_j)$.
	
	We define $\phi_i^u(b_j)$ as the closest vertex to $u$ in
	$C_{b_j}$ with label $i$ (with respect to~$\lab$), and $\rho_{i}^u(b_j) = \bm\rho_{i}^u(b_j)$, which is defined in \Cref{sec:DPboundedTW} according to the type of bag $b$. This way,~$\phi_i^u(b_j)$ automatically satisfies Property~\ref{prop:Sp1'} for $(\lab, \anch)$ restricted to $C_{b_j}$. In order to prove that the restriction of $(\lab, \anch)$ belongs to $\mathcal P(b_j,\phi(b_j),\rho(b_j)$, it therefore remains to show that this
	\lap also respects Properties~\ref{prop:Sp2'} and \ref{prop:Sp3'} regarding $b_j,\phi(b_j),\rho(b_j)$.
	
	Once all three requirements, $\phi_i^u(b_j)\in\bm \Phi_i^u(b_j)$, Property~\ref{prop:Sp2'} and Property~\ref{prop:Sp3'}, are verified, the definition for each bag type of $\bar A[b,\rho(b),\phi(b)]$ (\Cref{eq:twjoin,eq:twintro,eq:twforget}) and induction hypothesis on the children of $b$ lead to the desired
	inequality $A[b,\rho(b),\phi(b)] \geq \bar A[b,\rho(b),\phi(b)]$.\\
		
	\begin{itemize}\setlength\itemsep{6 pt}\setlength\parindent{12pt}
		\item \emph{Join nodes:} For a join node~$b$ with children $b_1,b_2$, we focus on a single child $b_j$. We first show that each $\phi_i^u(b_j)$ belongs to $\bm\Phi_i^u(b_j)$. 
		If $\phi_i^u(b)\in C_{b_j}$, then $\phi_i^u(b_j)=\phi_i^u(b)$ as desired. Otherwise, $\phi_i^u(b_j)$ cannot be closer to $u$ than $\phi_i^u(b)$, satisfying Condition~($\star$) from the definition of $\bm\Phi_i^u(b_j)$.
		
		Regarding Property~\ref{prop:Sp2'}, consider a vertex $v_1 \in C_{b_j}$ anchored to $v_2 = \anch(v_1)\notin C_{b_j}$ and a vertex $u\in S_b$ on the shortest path from $v_1$ to $v_2$. The objective is to show that $v_2 = \rho_{\lab(v_1)-1}^u(b_j)$.
		As $\anch$ is an anchoring of minimum cost that respects Property~\ref{prop:Sp2'}, we must have $v_2= \clo_u\{\rho_{\lab(v_1)-1}^u(b),\phi_{\lab(v_1)-1}^u(b) \}$. If $\phi_{\lab(v_1)-1}^u(b)\in C_{b_j}$, we must have $v_2= \rho_{\lab(v_1)-1}^u(b)$, as $v_2\notin C_{b_j}$. By definition of  $\phi_{\lab(v_1)-1}^u(b)$ and $\bm\rho_{\lab(v_1)-1}^u(b_j)$, we obtain $\bm\rho_{\lab(v_1)-1}^u(b_j) =\rho_{\lab(v_1)-1}^u(b)=v_2$.
		If $\phi_{\lab(v_1)-1}^u(b)\notin C_{b_j}$, then by definition $ v_2 = \bm \rho_{\lab(v_1)-1}^u(b_j)$. 
		
		For Property~\ref{prop:Sp3'}, consider $i$ and $u,v\in S_{b_j}$. Since~$(\lab, \anch) \in \mc P(b,\rho(b),\phi(b))$, we have $\dist{u,\rho_{i}^u(b)}\leq \dist{u,\rho_{i}^w(b)}$. From the definition of $\mc\rho_i(b)$, we get $\dist{u,\rho_{i}^u(b_j)}\leq \dist{u,\rho_{i}^w(b_j)}$.
		
		\item \emph{Forget nodes:} Recall that for a forget node $b$ with respect to $v$, we have $S_{b_1} = S_b\cup \{v\}$ and $C_{b_1} = C_b\setminus\{v\}$.  	
		Let $i_v = \lab(v)$, {which can be $\emptylabel$}. {Using the fact that $\phi(b)$ satisfies Property~\ref{prop:Sp1'}, it is easy to see that $i_v\in \bm I_v$.} %
		Clearly, the cost to anchor $v$ in $\lab$ equals~$c_{i_v}$.
		
		For a given $i$ and $u\in S_b$ (implying $u\neq v$), assume first that $\phi_i^u(b)\neq v$. Then, $\phi_i^u(b)\in C_{b_1}$ so we must have $\phi_i^u(b_1) = \phi_i^u(b) \in \bm\Phi_i^u(b_1)$. If on the other hand $\phi_i^u(b)= v$, then we must have $\dist{u,v} \leq \dist{u,\phi_i^u(b_1)}$, since $\anch\in P(b,\rho(b),\phi(b))$, so $\phi_i^u(b_1)\in \bm\Phi_i^u(b_1)$.
		
		We now want to show Property~\ref{prop:Sp2'}.
		Consider a vertex $v_1 \in C_{b_1}$ with label $\lab(v_1)$ anchored to $\anch(v_1)=v_2\notin C_{b_1}$. If $v_2=v$, then $v_2=\rho_{i_v}^v(b_1)$ so the property holds for this case. 	
		If $v_2\neq v$, then $v_2\notin C_b$, so there exists $u\in S_b$ such that $v_2=\rho_{\lab(v_1)-1}^u(b)$ as $\anch$ belongs to $\mc P(b,\rho(b),\phi(b))$. By definition, if $\rho_{\lab(v_1)-1}^u(b)\neq \rho_{\lab(v_1)-1}^u(b_1)$, then $\rho_{\lab(v_1)-1}^u(b_1) = v$ and $\dist{u,v} < \dist{u,\rho_{\lab(v_1)-1}^u(b_1)}$, which contradicts $\anch$ being a minimum cost anchoring.
		
		Property~\ref{prop:Sp3'} follows from the definition of $\mc \rho(b_1)$ and from the fact $(\lap,\anch)$ belongs to~$\mc P(b,\rho(b),\phi(b))$.

		We thus obtain $A[b, \rho(b), \phi(b)] = c_{i_v} +  A[b_1, \rho(b_1), \phi(b_1)]$, which proves the inequality.
		
		\item \emph{Introduce nodes:} If $b$ is an introduce node with respect to $v$, it has one child~$b_1$ with $S_{b_1} = S_b\setminus \{v\}$ and $C_{b_1} = C_b$.
		Since $(\lab, \anch)\in P(b,\rho(b),\phi(b))$, %
		Property~\ref{prop:Sp1'} implies that for all $i$, we have $\phi_i^v(b) = \clo_v\{\phi_i^u(b)\mid u\in S_{b_1}\}$.
		
		Thus, the conditions that would cause $\bar A[b,\rho(b),\phi(b)]$ to be infinite are not met.
		
		Consider a vertex $v_1 \in C_{b_1}$ with label $\lab(v_1)$ anchored to $\anch(v_1)=v_2\notin C_{b_1}$. There exists $u\in S_b$ such that $v_2=\rho_{\lab(v_1)-1}^u(b)$, since $\anch$ belongs to $\mc P(b,\rho(b),\phi(b))$. If $u\neq v$, we have the result. If $u=v$, then there exists a vertex $w\in S_{b_1}$ on the shortest path from $v_1$ to $v$. By Property~\ref{prop:Sp3'} for $\rho(b)$, we know that $\dist{w, \rho_{\lab(v_1)-1}^w(b)}\leq \dist{w, \rho_{\lab(v_1)-1}^v(b)}$. As $\anch$ is a minimum cost anchoring, this inequality must be an equality, so we obtain Property~\ref{prop:Sp2'}.
		
		Property~\ref{prop:Sp3'} holds for $\rho(b_j)$ as it is a subset of $\rho(b)$.
	\end{itemize}\vspace{1 pt}

	\begin{claim}\label{claim:TW_IEQ2}
		{We have $A[b,\rho(b),\phi(b)] \leq \bar A[b,\rho(b),\phi(b)]$.}
	\end{claim}
	Consider a bag $b$ and values $\rho(b)$ and $\phi(b)$. If $\bar A[b,\rho(b),\phi(b)]$ is infinite, the inequality trivially holds. We therefore consider the remaining cases. In particular, Property~\ref{prop:Sp3'} is respected regarding~$\rho(b)$, and
	the bag $b$ has either one child $b_1$ or two children $b_1$ and $b_2$. 
	Consider, for $j=1$ or $j\in\{1,2\}$, the anchoring guarantees $\bm \rho^u_i(b_j)$ and $\phi^u_i(b_j) \in \bm\Phi^u_i(b_j)$  yielding the
		value $\bar A[b,\rho(b),\phi(b)]$. In particular, $\bar A[b,\rho(b),\phi(b)]$ depends on the value(s) $\bar A[b_j,\bm\rho(b_j),\phi(b_j)]$. By the induction hypothesis, we have $\bar A[b_j,\bm\rho(b_j),\phi(b_j)]= A[b_j,\bm\rho(b_j),\phi(b_j)]$ so there exists a partial LAP $(\lab_j,\anch_j) \in \mc P(b_j,\rho(b_j),\phi(b_j))$ on each $C_{b_j}$ of cost $\bar A[b_j,\bm\rho(b_j),\phi(b_j)]$.

	Define $(\lab, \anch)$ to be the union of these {\lap}s in case of a join node. If $b$ is a forget node, then extend $(\lab, \anch)$ to $v$, by choosing the label $\lab(v)=i_v\in\bm I_v$ {(possibly $\emptylabel$)} which minimizes $\bar A[b,\rho(b),\phi(b)]$, as well as $\anch(v)=v$ if $i_v=\emptylabel$ and $\anch(v) = \clo_v\{\phi_{i_v-1}^v(b_1),\rho_{i_v-1}^v(b_1)\}$ otherwise. 
	And lastly, if $b$ is a introduce node, then simply keep the \lap since $C_b = C_{b_1}$.
	Thus, the cost of $(\lab,\anch)$ is precisely $\bar A[b,\rho(b),\phi(b)]$. We show
	that in all cases, the \lap~$(\lab, \anch)$  belongs to $\mc P(b,\rho(b),\phi(b))$. Then, by definition of
	$A[b,\rho(b),\phi(b)]$, the inequality holds. \\

	\begin{itemize}\setlength\itemsep{6 pt}\setlength\parindent{12pt}	
		\item \emph{Join node:} Consider a join node~$b$ with children $b_1,b_2$.  
		Since $C_{b_1}\cap C_{b_2}=\emptyset$, the union of the {\lap}s is well defined.
		
		Consider the anchoring guarantee $\phi_i^u(b)$, for some $u\in S_b$. We want to show that $\lab(\phi_i^u(b))=i$ and that no vertex of $C_b$ with label $i$ is closer to $u$. If $\phi_i^u(b)\in C_{b_1}$, then by definition, $\phi_i^u(b_1) = \phi_i^u(b)$. 
		Therefore, $\lab(\phi_i^u(b))=i$ and no vertex in $C_{b_1}$ closer to $u$ is labeled $i$. We also know {that $\dist{u,\phi_i^u(b)} \leq \dist{u,\phi_i^u(b_2)}$, by Condition ($\star$) in the definition of $\bm\Phi_i^u(b_1)$. Therefore, by symmetry, Property~\ref{prop:Sp1'} holds for $\phi_i^u(b)\in C_{b_2}$ as well. }

		{Regarding Property~\ref{prop:Sp2'}, consider a vertex $v_1$ of either of the $C_{b_j}$ anchored to a vertex $v_2 = \anch(v_1)\notin C_{b}$. Then, there exists some $u\in S_b$ such that $v_2 = \rho_{\lab(v_1)-1}^u(b_j)$ by definition of $\mc P(b_j,\rho(b_j), \phi(b_j))$, and since $v_2\notin C_b$, we must have $v_2 = \rho_{\lab(v_1)-1}^u(b)$.}

		\item \emph{Forget node:} Let $b$ be a forget node with respect to $v$. That is $S_{b_1} = S_b\cup \{v\}$ and $C_{b_1} = C_b\setminus\{v\}$. 
		Clearly, $(\lab, \anch)$ is well defined.
		
		{
			Consider $i$ and $u\in S_b$ (implying $u\neq v$), and assume first that $\phi_i^u(b) \neq v$. Then, we have $\phi_i^u(b) = \phi_i^u(b_1)$, so $\phi_i^u(b)$ is the closest vertex to $u$ with label $i$ in $C_{b_1}$. In order to show Property~\ref{prop:Sp1'} for this case, it remains to see that, if $i=i_v$, we have $\dist{u,v}\geq \dist{u,\phi_i^u(b)}$. This directly follows from the definition of $\bm I_v$. Assume now that  $\phi_i^u(b) = v$. By the definition of $\bm\Phi_i^u(b_1)$, we have $\dist{u,\phi_i^u(b_1)} \geq \dist{u,v}$. Then, by the definition of $\mc P(b_j,\rho(b_j),\phi(b_j))$, there is no vertex in $C_{b_1}$ with label $i$ closer to $u$ than $v$. This implies that $\phi_i^u(b) = v$ is the closest vertex to $u$ with label $i$ in $C_b$.
		}

		We now prove Property~\ref{prop:Sp2'}.
		Consider a vertex $v_1$ of $C_{b_1}$ (so $v_1\neq v$) with label $\lab(v_1)$ anchored to $v_2 =\anch(v_1)\notin C_{b}$. There exists some $u\in S_{b_1}$ such that $v_2 = \rho_{\lab(v_1)-1}^u(b_1)$ by definition of $\mc P(b_1,\rho(b_1), \phi(b_1))$ and $v_2\notin C_b$ so $v_2\neq v$.  Therefore, by definition of $\bm \rho_{\lab(v_1)-1}^u(b_1)$, there must exist some $w\in S_b$ such that $v_2 = \rho_{\lab(v_1)-1}^w(b)$.
		{Consider now $v$. If  $i_v=\emptylabel$ (which can only be the case if $v\notin \Term)$, we defined its anchor to be $\alpha(v)=v$.} If $i_v\neq\emptylabel$, we defined $\anch(v) = \clo_v\{\phi_{i_v-1}^v(b_1),\rho_{i_v-1}^v(b_1)\}$. If $\anch(v)\notin C_b$, then $\anch(v) = \clo_v\{\rho_{i_v-1}^u(b)\mid u\in S_b\}$ by definition of $\bm \rho_{i_v-1}^v(b_1)$, which completes the proof of Property~\ref{prop:Sp2'}.
		
		\item \emph{Introduce node:} Consider an introduce node $b$ with respect to $v$. That is $b$ has one child $b_1$ such that $S_{b_1} = S_b\setminus \{v\}$, $C_{b_1} = C_b$.	
		Again, $(\lab, \anch)$ is  obviously well defined.
		For each $i<k$ and $u\in S_{b_1}$, we have $\phi_i^u(b) = \phi_i^u(b_1)$, and we have $\phi_i^v(b)\neq \clo_v\{\phi_i^u(b)\mid u\in S_{b_1}\}$. As any path between $v$ and a vertex in $S_b$ contains a vertex in $S_{b_1}$, Property~\ref{prop:Sp1'} is satisfied.
		
		Consider a vertex $v_1$ of $C_{b}$ with label $\lab(v_1)$ anchored to $v_2 =\anch(v_1)\notin C_{b}$.
		There exists some $u\in S_{b_1}$ with $v_2 = \rho_{\lab(v_1)-1}^u(b_1)$ by definition of $\mc P(b_1,\rho(b_1), \phi(b_1))$. Since $S_{b_1}\subset S_b$, Property~\ref{prop:Sp2'} holds.
	\end{itemize}
	Thus, for all types of bags, $(\lab, \anch)$ is a member of $\mc P(b,\rho(b),\phi(b))$.
	Since $A[b,\rho(b),\phi(b)]$
	gives the minimum cost of all such {\lap}s, the claim %
	follows.
	\end{proof}

	\section{The \texorpdfstring{\kST}{k-hop M\v{S}T} Problem with Relaxed Hop Constraints}
	\label{sec:highway}
	
	In this section, we consider {the \kST problem in the relaxed model where {an algorithm may output a Steiner tree that uses more than $k$ hops while still} %
	comparing its performance to the optimal \kST}. On the positive side, we look at metrics of bounded highway dimension and present an M\v{S}T of near-optimal cost that violates the hop constraint by at most one hop. We further show that our reasoning yields an analogous results for metrics of bounded doubling dimension. 
	Note that this is by no means {a given}
	as Abraham et al.~\cite{AbrahamDFGW16} showed that constant doubling dimension does not imply constant highway dimension (whereas the converse is true). %
	{We also comment on the relaxation of the hop constraint of \kST in the case of general distance functions {induced by some} graph, which are not guaranteed to correspond to a metric. %
	{This setting does} not admit a polynomial time constant-factor approximation for \kST even with all distances being equal to $1$. Extending the hardness reduction from \cite{Manyem1996}, we show that the problem does not admit a constant-factor approximation even when the Steiner tree can use $k+{\ell}$ hops, for a constant ${\ell}$.}

	\subsection{Bounded Highway Dimension}
	{Feldmann et al.~\cite{highway} defined the highway dimension of a graph as follows. Let $B_r(v)=\{u\in V~|~\dist{u,v}\leq r\}$.} Given a universal constant $c>4$, the \emph{highway dimension} of a graph $G$ is the smallest integer $h$ such that for every $r\geq0$ and $v\in V$, there is a set of $h$ vertices in~$B_{cr}(v)$ that hits all shortest paths of length more than $r$ that lie entirely in~$B_{cr}(v)$.
	Before stating the results, we first define a $\delta$-net of a graph, which is informally a subset of vertices which are far from each other, while
	every vertex in the graph is close to this subset.
	Formally, a~$\delta$-net of a graph $G$, is a subset $U$ of $V$ such that for all
	$u\in V$, there exists $v\in U$ with $\dist{u,v}\leq \delta$ and for
	all $u,v\in U$, we have $\dist{u,v}>\delta$.
	
{Note that in the literature, there are definitions of highway dimensions different from the one above, both more general~\cite{AbrahamDFGW11} and more restricted~\cite{AbrahamDFGW16}.
	For further discussion, see~\cite{highway}.}

      {%
        Allowing the use of an additional hop, we can extend our algorithm for metrics {with} bounded treewidth to metrics of bounded highway dimensions and obtain the following result.} %
      
	\begin{theorem}\label{th:highway}
		For a metric induced by a graph of bounded highway dimension and a constant~$k$, let \optk be the cost of a \kST. A $(k+1)$-hop Steiner tree of cost at most $(1+\varepsilon)\optk$, for $\varepsilon >0$, can be computed in quasi-polynomial~time.
	\end{theorem}
		
	Feldmann et al.~\cite{highway} prove the following theorem, which gives sufficient conditions for a problem to admit a $QPTAS$ on graphs of constant highway dimension. {(We use the generic form of the theorem, which is not explicitly stated in~\cite{highway} but fully argued in the text.)} 
	
	\begin{theorem}[Reformulation of~{\cite[Theorem 8.1]{highway}}]
		\label{th:highwayreformulated}
		For a graph $G$ of constant highway dimension and a problem $\mc P$ satisfying conditions 1-6 below, a $(1+\varepsilon)$-approximation can be computed in quasi-polynomial time.
		\begin{enumerate}[nosep]
			\item An optimum solution of $\mc P$ can be computed in time $n^{O(\omega)}$ for graphs {with} treewidth $\omega$;
			\item A constant-approximation of $\mc P$ on metric graphs can be computed in polynomial time;
			\item The diameter of the graph can be assumed to be $O(n\cdot \opt_G)$, where $\opt_G$ is the cost of an optimal solution in $G$;
			\item An optimum solution for $\mc P$ on a $\delta$-net $U$ has cost at most $\opt_G + O(n\delta)$;
			\item The objective function of $\mc P$ is linear in the edge cost;
			\item A solution for $\mc P$ on a $\delta$-net $U$ can be converted to a solution on $V$ for {an additional} cost of $O(n\delta)$.
		\end{enumerate}
	\end{theorem}
	We now show that our main result, \Cref{thm:treewidth}, together with a previously known result, leads to \Cref{th:highway}, by using a slight variation of \Cref{th:highwayreformulated} to allow {the extra hop}.
	
	\begin{proof}[{Proof of}~\Cref{th:highway}]
		Applying \Cref{th:highwayreformulated}, it remains to verify that the \kST problem
		satisfies its six conditions, for $k$ constant, if we allow the algorithm to use one more hop (i.e.,
		computing a $(k+1)$-hop Steiner tree) than the optimal solution of cost $\optk$ to which we compare it.  The conditions and the explanation of why they are fulfilled are detailed below.
		\begin{enumerate}[nosep]
			\item {\Cref{thm:treewidth} states precisely this condition for \kST.}
			\item {For \kST in metric graphs, {Kantor and Peleg~\cite{KantorP09} presented} an algorithm with approximation factor $(1.52\!\cdot\! 9^{k-2})$.}
			\item 
			{The diameter of $G$ can be assumed to be $O(n\cdot \optk)$ since} edges of cost larger than $1.52\cdot9^{k-2}\cdot\optk$ can be deleted after computing the approximation of $\optk$ from~\cite{KantorP09}.
			\item 
			Consider an optimum \kST on $V$ and move each vertex not in $U$ to the closest vertex in $U$. This induces an extra cost of $O(n\delta)$ and is a solution on $U$.
			\item The objective function of \kST is indeed linear in the edge cost.
			\item This condition requires an additional hop. We claim that a solution for \kST on a $\delta$-net $U$ can be converted to a $(k+1)$-hop Steiner tree on $V$ for an additional cost of $O(n\delta)$. 
			Indeed, given a $k$-hop Steiner tree on $U$, we can anchor all vertices from $V\setminus U$ to their closest vertex in $U$ for an additional cost of $O(n\delta)$ and obtain a $(k+1)$-hop Steiner tree.	 
			This procedure of extending a solution is performed exactly once in the underlying algorithm. Therefore we can allow the algorithm to use one more hop on $G$ than the solution on $U$. Note that this property is not stated explicitly in~\cite{highway}.  
		\end{enumerate}
              \end{proof}

 	\subsection{Bounded Doubling Dimension}
 	The \emph{doubling dimension} $d$ of a graph $G$ refers to the smallest
	integer~$d$ such that any ball $B_{2r}(v)$ of radius $2r$ is contained in the union of~$2d$ balls of radius~$r$.

        {To show \Cref{th:highwayreformulated}, Feldman et al.~\cite{highway} construct a probabilistic embedding of metrics of bounded highway dimension into metrics {with} bounded treewidth that maintains distances approximately in expectation.
        They build upon the work of Talwar~\cite{Talwar04}, who gives an analogous result for metrics of bounded doubling dimension, and use it as a building block for their embedding algorithm.
        However, as discussed in \Cref{sec:results} and~\cite{highway}, the two dimensional parameters are not directly linked.}
        
	{The proof of~\Cref{th:highwayreformulated} fundamentally consists of replacing the given metric space by an appropriately chosen $\delta$-net, to limit the aspect ratio of the instance, and then using an algorithm for bounded treewidth graphs on the corresponding probabilistic embedding.
		Thus, we may use Talwar's embedding instead and replace ``highway dimension'' by ``doubling dimension'' in the statement of \Cref{th:highwayreformulated}. This yields the following.}

      \begin{theorem}
		For a metric of bounded doubling dimension and a constant $k$, let \optk be the cost of a \kST. A $(k+1)$-hop Steiner tree of cost at most $(1+\varepsilon)\optk$, for $\varepsilon >0$, can be computed in quasi-polynomial~time.
	\end{theorem}

       \subsection{{Non-Metric Distances}}
       {We have shown that relaxing the hop constraint only by {one} hop leads to efficient algorithms for some rather general metrics, such as {metrics with} bounded highway dimension {or bounded doubling dimension}.

        {We contrast these results by showing} that {even permitting} %
        {$\ell$ additional} hops{, for any constant $\ell$,} is not enough for the most gener{al} setting, which %
        {comprises} non-metric graphs with no constraints on the non-negative edge weights.
        {Specifically, the distance between two vertices may be infinite, so a direct connection may not exist.}
        {We generalize a} construction by Manyem and Stallmann~\cite{Manyem1996} {for showing an inapproximability result for \kST} and prove the following  result for \kST with relaxed hop constraints.}

\tikzstyle{helpervertex} = [draw=black,fill=white,circle,scale=0.5]

\begin{figure}[tb]
\centering
\begin{tikzpicture}[xscale = .59, yscale = .59]
  \node [root, label=$r$](r) at (0,10){};
  \node [vertex, label=right:$y_1$](y1) at ({-6},{4}){};
  \node [vertex, label=right:$y_2$](y2) at ({-6*(2-sqrt(3))}, {4}){};
  \node [vertex, label=right:$y_3$](y3) at ({6*(2-sqrt(3))}, {4}){};
  \node [vertex, label=right:$y_4$](y4) at (6,4){};

  \node [terminal, label=below:$x_1$](x1) at (-6.5,-0.5){};
  \node [terminal, label=below:$x_2$](x2) at (-3.5,-0.5){};
  \node [terminal, label=below:$x_3$](x3) at (0.5,-0.5){};
  \node [terminal, label=below:$x_4$](x4) at (3.5,-0.5){};
  \node [terminal, label=below:$x_5$](x5) at (6.5,-0.5){};

  \draw[black, thick](r) -- (y1);
  \draw[black, thick](r) -- (y2);
  \draw[black, thick](r) -- (y3);
  \draw[black, thick](r) -- (y4);

  \draw[black, thick, dashed](x1) -- (y1);
  \draw[black, thick, dashed](x2) -- (y1);
  \draw[black, thick, dashed](x4) -- (y1);

  \draw[black, thick, dashed](x1) -- (y2);
  \draw[black, thick, dashed](x3) -- (y2);
  \draw[black, thick, dashed](x4) -- (y2);

  \draw[black, thick, dashed](x3) -- (y3);
  \draw[black, thick, dashed](x5) -- (y3);

  \draw[black, thick, dashed](x4) -- (y4);
  \draw[black, thick, dashed](x5) -- (y4);

  \node [helpervertex](ry11) at ({-1.5}, {8.5}){};
  \node [helpervertex](ry12) at ({-3)}, {7}){};
  \node [helpervertex](ry13) at ({-4.5)}, {5.5}){};

  \node [helpervertex](ry21) at ({-1.5*(2-sqrt(3))}, {8.5}){};
  \node [helpervertex](ry22) at ({-3*(2-sqrt(3))}, {7}){};
  \node [helpervertex](ry23) at ({-4.5*(2-sqrt(3))}, {5.5}){};

  \node [helpervertex](ry31) at ({1.5*(2-sqrt(3))}, {8.5}){};
  \node [helpervertex](ry32) at ({3*(2-sqrt(3))}, {7}){};
  \node [helpervertex](ry33) at ({4.5*(2-sqrt(3))}, {5.5}){};

  \node [helpervertex](ry41) at ({1.5}, {8.5}){};
  \node [helpervertex](ry42) at ({3)}, {7}){};
  \node [helpervertex](ry43) at ({4.5)}, {5.5}){};

  \node [helpervertex](x1y11) at ($(x1)!0.33333!(y1)$) {};
  \node [helpervertex](x1y12) at ($(x1)!0.66666!(y1)$) {};
  
  \node [helpervertex](x5y41) at ($(x5)!0.33333!(y4)$) {};
  \node [helpervertex](x5y42) at ($(x5)!0.66666!(y4)$) {};
  
  \node [helpervertex](x2y11) at (intersection of x1y11--x5y41 and x2--y1) {};
  \node [helpervertex](x2y12) at (intersection of x1y12--x5y42 and x2--y1) {};
  
  \node [helpervertex](x4y11) at (intersection of x1y11--x5y41 and x4--y1) {};
  \node [helpervertex](x4y12) at (intersection of x1y12--x5y42 and x4--y1) {};

  \node [helpervertex](x4y11) at (intersection of x1y11--x5y41 and x4--y1) {};
  \node [helpervertex](x4y12) at (intersection of x1y12--x5y42 and x4--y1) {};

  \node [helpervertex](x1y21) at (intersection of x1y11--x5y41 and x1--y2) {};
  \node [helpervertex](x1y22) at (intersection of x1y12--x5y42 and x1--y2) {};

  \node [helpervertex](x3y21) at (intersection of x1y11--x5y41 and x3--y2) {};
  \node [helpervertex](x3y22) at (intersection of x1y12--x5y42 and x3--y2) {};

  \node [helpervertex](x4y21) at (intersection of x1y11--x5y41 and x4--y2) {};
  \node [helpervertex](x4y22) at (intersection of x1y12--x5y42 and x4--y2) {};

  \node [helpervertex](x3y31) at (intersection of x1y11--x5y41 and x3--y3) {};
  \node [helpervertex](x3y32) at (intersection of x1y12--x5y42 and x3--y3) {};

  \node [helpervertex](x5y31) at (intersection of x1y11--x5y41 and x5--y3) {};
  \node [helpervertex](x5y32) at (intersection of x1y12--x5y42 and x5--y3) {};

  \node [helpervertex](x4y41) at (intersection of x1y11--x5y41 and x4--y4) {};
  \node [helpervertex](x4y42) at (intersection of x1y12--x5y42 and x4--y4) {};

 \draw[decoration={calligraphic brace,amplitude=5pt,mirror}, decorate, line width=1.25pt] (7.5,-0.5) node{} -- (7.5,4) node[midway,right,xshift=3]{$\ell=3$}; 
 \draw[decoration={calligraphic brace,amplitude=5pt,mirror}, decorate, line width=1.25pt] (7.5,4.2) node{} -- (7.5,10) node[midway,right,xshift=3]{$t=4$}; 

\end{tikzpicture}
\caption{Reduction from \textsc{Unweighted Set Cover} to \kST with relaxed hop constraints. The corresponding set cover universe is $X = \{x_1, \ldots, x_5\}$ with sets $Y_1 = \{x_1, x_2, x_4\}$, $Y_2 = \{x_1, x_3, x_4\}$, $Y_3 = \{x_3, x_5\}$ and $Y_4 = \{x_4, x_5\}$. The vertices $x_1, \ldots, x_5$ are set to be terminals. All edges in the graph have length $1$.}

\label{fig:additive-hops-hardness}
\end{figure}

          \begin{theorem}
          	{For any constant $c$ and $\ell$, and given a {weighted} graph $G$ %
          	as input, it is NP-hard to find a $(k+\ell)$-hop Steiner tree in $G$ of cost $(1-c)\cdot\log n \cdot OPT$, where $OPT$ is the {minimal} cost of a $k$-hop Steiner tree in $G$.
          	The above statement is true even when all edges in the graph $G$ have weight equal to $1$.}
          \end{theorem}
          \begin{proof}
          	{We reduce from \textsc{Unweighted Set Cover}. There, we are given a family of sets $\mathcal{Y} = \{Y_1, \ldots, Y_m\}$ over a shared universe $X = \{x_1, \ldots, x_n\}$ and we are tasked with deciding whether there is a selection $\mathcal{S} \subseteq \mathcal{Y}$
          		of {at most} $s$ sets such that $\bigcup_{Y \in \mathcal{S}} Y = X$. We start our reduction by representing the problem {as} a bipartite graph, with edges $\{x_i,y_j\}$ corresponding to $x_i \in Y_j$. Next, we add a new vertex $r$ to serve as our root, and we set
          		every vertex $x_i$ to be a terminal of the Steiner tree. We connect $r$ to each vertex $y_j$ by a unique path of $t$ edges of weight $1$ each. Finally, we subdivide each edge $\{x_i, y_j\}$ to form a path of $\ell$ edges, again each single edge having weight exactly $1$. See \Cref{fig:additive-hops-hardness} for an example.}
          	          	
          	{{Let $k=t+\ell$.} We first observe that if there is a solution $\mathcal{S}$ to the \textsc{Unweighted Set Cover} problem of size $s$, then there exists a $k$-hop Steiner tree of total {cost $st + n\ell$}. %
          		Such a tree can be created by %
          		including the $y_j$ corresponding to $Y_j \in \mathcal{S}$ in the Steiner tree, and connecting any $x_i$ to {exactly} one of the $y_j$ for which $x_i \in Y_j \in \mathcal{S}$ {paying a cost of $\ell$ each}. Finally, %
          		{connect the selected $y_j$ via the direct paths of cost $t$ to the root $r$.}}
          	
          	{Next, we consider any $k+({2}\ell-1)$-hop Steiner tree $T$ on this instance and a node $y_j$ that is included in the Steiner tree. In this case, $y_j$ must be connected directly to the root by the path of weight $t$. This is true as the shortest hop distance between any two vertices $y_j, y_{j'}$ is at least $2\ell$. Since the number of hops to connect some $y_{j'}$ to the root is at least $t$, connecting $y_j$ via $y_{j'}$ would contradict the assumption that we are given a $t+(2\ell-1)$-hop Steiner tree. It follows, that $st + n\ell$ indeed is also the minimum cost of a Steiner tree with at most $k+(2\ell-1)$ hops that contains $s$ of the $y_j$.}
          	
          	{With the above, %
          		we see that any Steiner tree with at least $k$ hops and at most $k+({2}\ell-1)$ hops directly corresponds to a solution $\mathcal{S}$ of the \textsc{Unweighted Set Cover} problem {that consists of exactly the sets $Y_j$ for which
          			vertex $y_j$ is included in the Steiner tree.} %
          		Furthermore, any such Steiner tree with at most $k+({2}\ell-1)$ hops {and} of cost at most {$st + n\ell$} %
          		corresponds to a set cover with at most {$s$} sets. }
          		
          	{Assume we can obtain a $k+\ell$-hop Steiner tree of cost at most $(1-c)\log(n)\cdot\opt.$
          		Denoting by $s^*$ the size of an minimum set cover, this equals $(1-c)\log(n)\cdot(ts^*+\ell n) = (1-\frac c 2)\log(n)\cdot ts^*+ (1-c)\log(n)\cdot\ell n - \frac c 2\log(n) \cdot ts^*$ which is at most $(1-\frac c 2)\log(n)\cdot ts^*+ \ell n$ when substituting $t = \left\lceil\frac{\ell n ((1-c)\log (n) -1)}{\frac{c}{2} s^* \log n} \right\rceil$, which is polynomial in $n$.
          		Thus, the Steiner tree corresponds to a set cover of size at most $(1-\frac c 2)\log(n)\cdot s^*$.
          		
          		Summing up, we gave a polynomial-size reduction that implies that given a $(1-c)\log(n)$ approximation for the \kST problem, we obtain a $(1-c')\log(n)$ approximation for set cover, with $c,c'>0$. The latter, however, is known to be \NP-hard~\cite{Moshkovitz12} concluding the proof.
          	}
          \end{proof}

        \section{Conclusions}
        {In this work, we show how to solve the \kST problem {exactly and efficiently} in tree-like metrics and {extend} our results for other metrics, such as bounded highway dimension and doubling metrics. {In our extensions, we have relaxed the constraint on the number of hops $k$ to $k+1$;} it remains open whether this {relaxation} is needed for the latter two classes. {While soft hop constraints, at least with an additive constraint, do not help in the non-metric setting,} it would be interesting to know {whether} {one can %
        obtain} a constant{-factor} approximation for the \kST problem in arbitrary metrics under a soft hop constraint, {or even an approximation scheme},
          which is ruled out for hard hop constraints \cite{Guha1999, Alfandari1999}.

Further, our exact algorithms raise the question whether {the problem of} finding a \kST in a metric induced by a tree is fixed-parameter tractable by the parameter $k$, i.e., does exist an algorithm with running time $O(\text{poly}(n)\cdot f(k))$ for some function $f$.

\section*{Acknowledgments} 
We thank Ji\v r\' i Sgall for fruitful {discussions on the \kMST problem in  path metrics.}

\bibliographystyle{siamplain}
\bibliography{khop}
\end{document}